\documentclass[letterpaper, 10 pt, conference]{ieeeconf}  % Comment this line out
\IEEEoverridecommandlockouts                              % This command is only
\usepackage{amsmath} % assumes amsmath package installed
\usepackage{amssymb}  % assumes amsmath package installed

\usepackage{cite} % group citations
\usepackage{tikz}
\usepackage{breqn} % automatic multi-line equations 
\usepackage{algorithm}
\usepackage{algpseudocode}
\algrenewcommand\algorithmicrequire{\textbf{Input:}}
\algrenewcommand\algorithmicensure{\textbf{Output:}}

\newtheorem{remark}{Remark}
\newtheorem{lemma}{Lemma}
\newtheorem{theorem}{Theorem}

\usepackage{etoolbox}
\makeatletter
\patchcmd{\@makecaption}
  {\scshape}
  {}
  {}
  {}
\makeatletter
\patchcmd{\@makecaption}
  {\\}
  {.\quad}
  {}
  {}
\makeatother

\title{\LARGE \bf
Scalable Forward Reachability Analysis of Multi-Agent Systems with Neural Network Controllers}

\author{Oliver Gates, Matthew Newton and Konstantinos Gatsis
\thanks{© 2023 IEEE.  Personal use of this material is permitted.  Permission from IEEE must be obtained for all other uses, in any current or future media, including reprinting/republishing this material for advertising or promotional purposes, creating new collective works, for resale or redistribution to servers or lists, or reuse of any copyrighted component of this work in other works.}%
\thanks{The authors are with the Department of Engineering Science, University of Oxford, UK. Email: oliver.gates@st-hildas.ox.ac.uk, matthew.newton@worc.ox.ac.uk, konstantinos.gatsis@eng.ox.ac.uk}}

\newcommand{\rowtwo}[2]{\begin{bmatrix}
#1 & #2
\end{bmatrix}}

\newcommand{\rowthree}[3]{\begin{bmatrix}
#1 & #2 & #3
\end{bmatrix}}

\newcommand{\rowfour}[4]{\begin{bmatrix}
#1 & #2 & #3 & #4
\end{bmatrix}}

\newcommand{\coltwo}[2]{\begin{bmatrix}
#1 \\
#2
\end{bmatrix}}

\newcommand{\colthree}[3]{\begin{bmatrix}
#1 \\
#2 \\
#3
\end{bmatrix}}

\newcommand{\mattwo}[4]{\begin{bmatrix}
#1 & #2 \\
#3 & #4
\end{bmatrix}}

\newcommand{\matthree}[9]{\begin{bmatrix}
#1 & #2 & #3 \\
#4 & #5 & #6 \\
#7 & #8 & #9
\end{bmatrix}}

\newcommand{\bigzero}{\mbox{\normalfont\Large 0}}

\usepackage{cuted}

\begin{document}

\maketitle
\thispagestyle{empty}
\pagestyle{empty}

%%%%%%%%%%%%%%%%%%%%%%%%%%%%%%%%%%%%%%%%%%%%%%%%%%%%%%%%%%%%%%%%%%%%%%%%%%%%%%%%
\begin{abstract}

Neural networks (NNs) have been shown to learn complex control laws successfully, often with performance advantages or decreased computational cost compared to alternative methods. Neural network controllers (NNCs) are, however, highly sensitive to disturbances and uncertainty, meaning that it can be challenging to make satisfactory robustness guarantees for systems with these controllers. This problem is exacerbated when considering multi-agent NN-controlled systems, as existing reachability methods often scale poorly for large systems. This paper addresses the problem of finding overapproximations of forward reachable sets for discrete-time uncertain multi-agent systems with distributed NNC architectures. We first reformulate the dynamics, making the system more amenable to reachablility analysis. Next, we take advantage of the distributed architecture to split the overall reachability problem into smaller problems, significantly reducing computation time. We use quadratic constraints, along with a convex representation of uncertainty in each agent's model, to form semidefinite programs, the solutions of which give overapproximations of forward reachable sets for each agent. Finally, the methodology is tested on two realistic examples: a platoon of vehicles and a power network system.

\end{abstract}

%%%%%%%%%%%%%%%%%%%%%%%%%%%%%%%%%%%%%%%%%%%%%%%%%%%%%%%%%%%%%%%%%%%%%%%%%%%%%%%%
\section{INTRODUCTION}

There has been recent interest in the use of neural networks (NNs) for control in closed-loop feedback systems. Neural network controllers (NNCs) can be used to imitate traditional control policies, such as model predictive control (MPC), with reduced computational cost \cite{122}, or to implement deep reinforcement learning (RL) policies \cite{138}. Even for simple linear systems, NNCs can be used to implement complex non-linear control laws (which may not be easy to achieve with existing methods). NNs are, however, highly sensitive to input perturbations, so disturbances in the closed-loop system can have adverse effects \cite{002}. This is problematic when NNCs are used in safety-critical systems, and recent work has focused on reachability analysis for systems with NNCs; if we can overapproximate the forward reachable sets, then it can be verified that certain regions of the state space are avoided over a given horizon.

The problem of computing forward reachable sets becomes more challenging when considering a multi-agent system in which each agent is controlled by an NN (or a series of NNs); the effects of small perturbations to the input of one NNC are propagated through the system. Control of multi-agent systems is well-studied, and common goals include consensus, formation control and flocking/swarming \cite{112}. Multi-agent control architectures can be categorised according to the dependence of each agent's control input on other agents' states: (a) centralised control, in which each agent's control input is a function of all agents' states; (b) distributed control, in which each agent's control input is a function of a subset of the other agents' states; (c) decentralised control, in which each agent's control input is a function of only its own state \cite{038}. In distributed control, common approaches include state feedback \cite{139} and distributed model predictive control (DMPC) \cite{097}.

A number of methods have been proposed for the reachability analysis of systems with NNCs \cite{007,132,032,042,034,142,140,149,070}. %\cite{007,132,032,042,034,142,140,050,051,070,079,080}.
In general, the problem of reachability for discrete-time LTI systems is undecidable \cite{141}, so methods for safety verification of closed-loop systems often aim to find tight overapproximations of the forward reachable sets. Alternatively, additional restrictions can be placed on the problem to allow the exact reachable sets to be computed. Generally, there is a tradeoff between scalability and tightness of the bounds \cite{002}.

In \cite{007}, semidefinite programming (SDP) is used to compute overapproximations of the forward reachable sets, and \cite{132} builds on this work by considering parameter-varying systems. In \cite{032} and \cite{042}, the input set is partitioned into smaller sets, and a linear programming (LP) approach is used to overapproximate the reachable sets; in \cite{032}, the input set partitioning approach is also applied to the method in \cite{007}, and a comparison is made between the LP and SDP approaches, demonstrating that the former is faster but the latter results in tighter bounds. The work in \cite{034} restricts the input sets to be constrained zonotopes, allowing for the exact computation of reachable sets, and the work in \cite{142} represents the input sets as hybrid zonotopes, allowing for a class of non-convex input sets. In \cite{140}, polynomial zonotopes are used to abstract the closed-loop dynamics, providing tight overapproximations. Other approaches include the use of polynomial optimisation \cite{149} and Bernstein polynomials \cite{070}.
% Other approaches include branch-and-bound-based methods \cite{050} \cite{051}, Bernstein polynomials \cite{070}, mixed monotone theory \cite{079} and sampling-based methods \cite{080}.

Existing reachability methods for NN-controlled systems do not explicitly consider multi-agent systems with distributed control architectures. Similarly to the single-agent case, in which NNCs can be used to learn complex control laws, a distributed NNC architecture can be used to learn complex distributed control laws \cite{148}. An example is DMPC, in which each agent's controller solves an optimisation problem based on its own state and those of its neighbours. An overview of DMPC is given in \cite{097}, and its applications include vehicle platooning \cite{022}, frequency regulation in power systems \cite{074} and formation control of UAVs \cite{143}.

%\subsection{Contributions}

In this paper, we present a scalable method to compute overapproximations of the forward reachable sets for uncertain multi-agent systems with distributed NNC architectures. To the best of our knowledge, this is the first work which explicitly deals with the multi-agent NNC reachability problem. The main contributions are as follows:
\begin{itemize}
    \item we reformulate the dynamics, making the system more amenable to reachability analysis;
    \item we take advantage of the distributed architecture to split the overall reachability problem into smaller problems, using an existing SDP-based approach to overapproximate the forward reachable sets for each agent, and further extend this result to incorporate model uncertainty in the agents' dynamics;
    \item we demonstrate the effectiveness of this methodology on two realistic multi-agent systems with different structures: a platoon of vehicles and a power network;
    \item we compare our approach to the approach of treating the multi-agent system as one overall system, and we demonstrate that our approach outperforms the alternative approach.
\end{itemize}
In Section \ref{sec:PS}, we describe the dynamics, give some remarks about the form of the controller and describe the forward reachability problem. In Section \ref{sec:reformDyn}, we provide a simplification of the dynamics and control input, and in Section \ref{sec:reach}, we present the reachability method. In Section \ref{sec:MU}, we introduce model uncertainty. In Section \ref{sec:experiments}, we present experiments to demonstrate the method.

\subsection{Notation}

The set of real $n\times m$ matrices is denoted by $\mathbb{R}^{n \times m}$, the set of real $n$-length vectors by $\mathbb{R}^n$ and the set of real numbers by $\mathbb{R}$. The set of symmetric $n \times n$ matrices is denoted by $\mathbb{S}^n$. The set of diagonal $n \times n$ matrices is denoted by $\mathbb{D}^n$. The set of positive integers is denoted by $\mathbb{Z}^+$. The cardinality of a set $\mathcal{S}$ is denoted by $|\mathcal{S}|$. The $n \times n$ identity matrix is denoted by $I_n$. The symbols $\geq$ and $\leq$ apply elementwise to vectors and matrices. $A \preceq 0$ implies that matrix $A$ is negative semidefinite. The number $0$ is used to represent the scalar, vector or matrix of appropriate size; the size will be clear from the context.

For clarity, whenever the letters $i$ and $j$ are used in this paper, they refer to the index of an agent.

%%%%%%%%%%%%%%%%%%%%%%%%%%%%%%%%%%%%%%%%%%%%%%%%%%%%%%%%%%%%%%%%%%%%%%%%%%%%%%%%
\section{PROBLEM STATEMENT}\label{sec:PS}

\subsection{Multi-agent dynamics}

We consider a discrete-time multi-agent system of $M$ agents, where each agent $i$ has linear time-invariant dynamics
\begin{equation}
    x^{[i]}_{k+1} = A_{ii}x^{[i]}_k + \sum_{j \in \mathcal{N}_i} A_{ij}x^{[j]}_k + B_iu^{[i]}_k + w^{[i]}_k, \label{eq:agentDynamics}
\end{equation}
where for each agent $i \in \{1,\dots,M\} = \mathcal{I}$, $x^{[i]}_k \in \mathbb{R}^{n_x}$ is the local state, $x^{[j]}_k \in \mathbb{R}^{n_x}$ is the $j^{\text{th}}$ neighbouring state, $u^{[i]}_k \in \mathbb{R}^{n_u}$ is the control input, $w^{[i]}_k \in \mathbb{R}^{n_x}$ is a known external input, $A_{ii} \in \mathbb{R}^{n_x \times n_x}$ is the state matrix, $A_{ij} \in \mathbb{R}^{n_x \times n_x}$ is the matrix describing the effect of state $x^{[j]}_k$ on agent $i$, $B_i \in \mathbb{R}^{n_x \times n_u}$ is the input matrix, and $\mathcal{N}_i$ is the set of neighbours. Note that for simplicity, we have assumed that all agents have the same dimensions (although this assumption could be relaxed). We also assume that $|\mathcal{N}_i|>0\ \forall i$.

In this paper, without loss of generality, we focus on distributed control architectures, in which each agent's control input is a function of a subset $\mathcal{N}_i$ of all other agents' states $\mathcal{I}$. The methods presented in this paper can be easily extended to the other two cases by setting $\mathcal{N}_i = \mathcal{I}$ (centralised) or $\mathcal{N}_i = \{i\}$ (decentralised).

\subsection{Control input}

In a traditional distributed control scheme with proportional feedback, the control input $u^{[i]}_{\mathrm{trad},k}$ might be given by
\begin{equation*}
    u^{[i]}_{\mathrm{trad},k} = \sum_{j \in \mathcal{N}_i} K_{ij}\left(x^{[j]}_k-x^{[i]}_k\right), %\label{eq:tradControlInput}
\end{equation*}
where $K_{ij} \in \mathbb{R}^{n_u \times n_x}$ is some gain matrix (which could represent multiple gain matrices) \cite{lewis2014cooperative}. Similarly, in a DMPC scheme, the $i^{\text{th}}$ control input is generated by solving an optimisation problem based on the agent's state $x^{[i]}_k$ and the neighbours' states $x^{[j]}_k\ \forall j \in \mathcal{N}_i$. In this paper, we consider the extension of the traditional distributed control schemes to NN-based control, in which the $i^{\text{th}}$ control input is generated by some non-linear function of the agent's state $x^{[i]}_k$ and the neighbours' states $x^{[j]}_k\ \forall j \in \mathcal{N}_i$. We also consider the possibility of controller saturation. Hence, the $i^{\text{th}}$ control input $u^{[i]}_k$ is given by
\begin{equation}
    u^{[i]}_k = \mathrm{sat}_{\mathcal{U}_i}\left[ \sum_{j \in \mathcal{N}_i} \pi_{ij}\left(\coltwo{x^{[i]}_k}{x^{[j]}_k}\right) \right], \label{eq:controlInput}
\end{equation}
where $\mathrm{sat}_{\mathcal{U}_i}$ is a projection onto the set $\mathcal{U}_i = \{ u \in \mathbb{R}^{n_u}\ |\ \underline{u}_i \leq u \leq \overline{u}_i\}$, where $\underline{u}_i$ and $\overline{u}_i$ are the lower and upper limits, respectively, for the $i^{\text{th}}$ controller, and $\pi_{ij}: \mathbb{R}^{2n_x} \rightarrow \mathbb{R}^{n_u}$ is a function representing the mapping of the input through a multi-layer perceptron (MLP).
% \begin{remark}
% In (\ref{eq:controlInput}), we consider the general case, in which the input to each NN $\pi_{ij}$ is the concatenation of the agent's state $x^{[i]}_k$ and the neighbour's state $x^{[j]}_k$, i.e. {\footnotesize$\rowtwo{{x^{[i]}_{k}}^{\top}}{{x^{[j]}_{k}}^{\top}}^{\top}$}. It is possible to directly consider more specific cases by instead considering the input to the MLP as a linear transformation of the aforementioned input, i.e. {\footnotesize$F_{ij}\rowtwo{{x^{[i]}_{k}}^{\top}}{{x^{[j]}_{k}}^{\top}}^{\top}$}, where $F_{ij} \in \mathbb{R}^{n_F \times 2n_x}$. For example, we could consider the direct extension of (\ref{eq:tradControlInput}), taking the input to the MLP as the difference between the neighbour's state and the agent's state, by using $F_{ij} = \rowtwo{-I_{n_x}}{I_{n_x}}$.
% \end{remark}
\begin{remark}
In (\ref{eq:controlInput}), we consider a separate NN $\pi_{ij}$ for each neighbouring agent, as this preserves the ability to consider the effects of each neighbour individually, by isolating the effect of a particular neighbour's contribution to the control input. However, the control input could also be generated by feeding the state of the agent and the states of its neighbours into a single MLP $\Pi_i : \mathbb{R}^{(1+|\mathcal{N}_i|)n_x} \rightarrow \mathbb{R}^{n_u}$. A transformation between the two architectures is given in Section \ref{subsec:inputTransform}.
\end{remark}

\subsection{Multi-layer perceptron}

The mapping $s \mapsto \pi_{ij}(s)$ for an $L$-layer MLP is
\begin{subequations}
\begin{align}
    z^0_{ij} &= s, \label{eq:MLP1} \\
    z^{\ell+1}_{ij} &= \sigma^{\ell}\left( W^{\ell}_{ij} z^{\ell}_{ij} + b^{\ell}_{ij} \right), \quad \ell = 0,\dots,L-1, \\
    \pi_{ij}(s) &= W^{L}_{ij} z^{L}_{ij} + b^{L}_{ij}, \label{eq:MLP3}
\end{align}
\end{subequations}
where $z^{\ell}_{ij} \in \mathbb{R}^{n_{\ell}}$ is the $\ell^{\text{th}}$ vector of activation values (note that $n_0 = 2n_x$), $W^{\ell}_{ij} \in \mathbb{R}^{n_{\ell+1} \times {n_{\ell}}}$ is the $\ell^{\text{th}}$ weight matrix, $b^{\ell}_{ij} \in \mathbb{R}^{n_{\ell+1}}$ is the $\ell^{\text{th}}$ bias vector, and $\sigma^{\ell} : \mathbb{R}^{n_{\ell+1}} \rightarrow \mathbb{R}^{n_{\ell+1}}$ is the $\ell^{\text{th}}$ ReLU activation function, i.e. $\sigma^{\ell}(y) = \max(y,0)$, applied elementwise, such that $\sigma^{\ell}(y) = \rowthree{\max(y_1,0)}{\cdots}{\max(y_{n_{\ell}+1},0)}^{\top}$,
% \begin{equation*}
%     \sigma^{\ell}(y) = \rowthree{\max(y_1,0)}{\cdots}{\max(y_{n_{\ell}+1},0)}^{\top},
% \end{equation*}
where $y =\rowthree{y_1}{\dots}{y_{n_{\ell}+1}}^{\top}$ is the vector of pre-activation values. For simplicity, we assume that all MLPs have the same structure (size and number of hidden layers) for all $i,j$.

\subsection{Overapproximation of forward reachable sets}

We denote the set of all possible states of the $i^{\text{th}}$ agent at time $k$ as $\mathcal{X}^{[i]}_k$, such that $x^{[i]}_k \in \mathcal{X}^{[i]}_k$. Given $\mathcal{X}^{[i]}_k$ and the sets of neighbouring states at time $k$, i.e. $\mathcal{X}^{[j]}_k\ \forall j \in \mathcal{N}_i$, we aim to find an overapproximation $\widehat{\mathcal{X}}^{[i]}_{k+1}$ of the reachable set $\mathcal{X}^{[i]}_{k+1}$ at the next time step. Specifically, we aim to find the tightest polytopic overapproximation of the reachable set
\begin{subequations}
    \begin{align}
        \min &\quad \mathrm{vol}\left(\widehat{\mathcal{X}}^{[i]}_{k+1}\right) \label{eq:originalProblemA}\\
        \text{subject to} &\quad \widehat{\mathcal{X}}^{[i]}_{k+1} \supseteq \mathcal{X}^{[i]}_{k+1},\label{eq:originalProblemB}\\ &\quad \widehat{\mathcal{X}}^{[i]}_{k+1} \ \text{is a polytope}, \label{eq:originalProblemC}
    \end{align} 
\end{subequations}
for each agent $i$, where $\mathrm{vol}$ is the $n_x$-dimensional volume. This could then be applied recursively to overapproximate the next forward reachable sets $\mathcal{X}^{[i]}_{k+2},\dots,\mathcal{X}^{[i]}_{k+N}$ over a finite horizon $N$, and used to verify that certain unsafe regions of the state space are avoided over this horizon.

%%%%%%%%%%%%%%%%%%%%%%%%%%%%%%%%%%%%%%%%%%%%%%%%%%%%%%%%%%%%%%%%%%%%%%%%%%%%%%%%
\section{REFORMULATION OF DYNAMICS} \label{sec:reformDyn}

An obvious approach to the forward reachability problem is to augment the agents' states into one `overall' state
\begin{equation*}
    x_k = \rowthree{{x^{[1]}_k}^{\top}}{\cdots}{{x^{[M]}_k}^{\top}}^{\top},
\end{equation*}
then form a recursion for the overall dynamics and use existing methods to perform reachability analysis on this system. The main issue with this approach is that it ignores the distributed architecture of the system -- the dynamics of each agent depend only on its neighbours, not on all other agents. As a result, we reformulate the dynamics given by (\ref{eq:agentDynamics}) and (\ref{eq:controlInput}) to allow reachability analysis to be performed for each agent. In Section \ref{sec:experiments}, we show that there is a significant computational advantage to solving $M$ smaller reachability problems over one large reachability problem for the SDP-based method. We note that the approach of decomposing the reachability problem into smaller problems has been used more generally in work on reachability analysis \cite{161,162}.

\subsection{Simplification of dynamics}

Let $g(i,j)$ be a function which returns the $j^{\text{th}}$ neighbour of the $i^{\text{th}}$ agent. For example, if agent $3$ has $\mathcal{N}_3 = \{2,5,6\}$, then $g(3,1) = 2$, $g(3,2) = 5$ and $g(3,3) = 6$. Then we can write $\mathcal{N}_i = \left\{g(i,1),\dots,g(i,q_i)\right\}$,
% \begin{equation*}
%     \mathcal{N}_i = \left\{g(i,1),\dots,g(i,q_i)\right\},
% \end{equation*}
where $q_i = |\mathcal{N}_i|$, and let
\begin{align}
    \tilde{A}_i &= \rowfour{A_{ii}}{A_{ig(i,1)}}{\cdots}{A_{ig(i,q_i)}}, \label{eq:Atilde} \\
    \tilde{x}^{[i]}_k &= \rowfour{{x^{[i]}_{k}}^{\top}}{{x^{[g(i,1)]}_{k}}^{\top}}{\cdots}{{x^{[g(i,q_i)]}_{k}}^{\top}}^{\top}, \label{eq:xtilde}
\end{align}
such that $\tilde{x}^{[i]}_k$ is the concatenation of the state of agent $i$ and the states of the neighbours of agent $i$ at time $k$, then (\ref{eq:agentDynamics}) can be written as
\begin{equation*}
    x^{[i]}_{k+1} = \tilde{A}_i\tilde{x}^{[i]}_k + B_iu^{[i]}_k + w^{[i]}_k.
\end{equation*}

\subsection{Simplification of control input}\label{subsec:inputTransform}

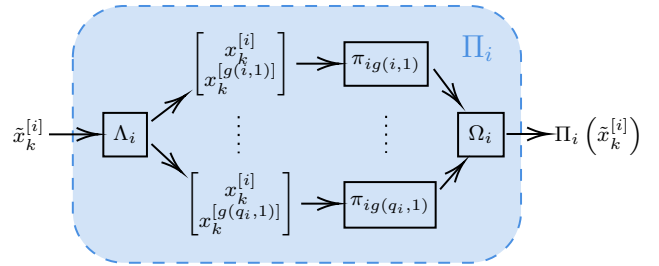
\begin{figure}[h]
  \centering

\tikzset{every picture/.style={line width=0.75pt}} %set default line width to 0.75pt        

\begin{tikzpicture}[x=0.75pt,y=0.75pt,yscale=-1,xscale=1]
%uncomment if require: \path (0,300); %set diagram left start at 0, and has height of 300

%Rounded Rect [id:dp03230882045914729] 
\draw  [color={rgb, 255:red, 74; green, 144; blue, 226 }  ,draw opacity=1 ][fill={rgb, 255:red, 74; green, 144; blue, 226 }  ,fill opacity=0.25 ][dash pattern={on 4.5pt off 4.5pt}] (226.27,98.11) .. controls (226.27,83.8) and (237.87,72.19) .. (252.19,72.19) -- (426.41,72.19) .. controls (440.73,72.19) and (452.33,83.8) .. (452.33,98.11) -- (452.33,175.88) .. controls (452.33,190.2) and (440.73,201.8) .. (426.41,201.8) -- (252.19,201.8) .. controls (237.87,201.8) and (226.27,190.2) .. (226.27,175.88) -- cycle ;
%Straight Lines [id:da021630083098125974] 
\draw    (214.25,136.99) -- (237.12,136.99) ;
\draw [shift={(239.12,136.99)}, rotate = 180] [color={rgb, 255:red, 0; green, 0; blue, 0 }  ][line width=0.75]    (10.93,-3.29) .. controls (6.95,-1.4) and (3.31,-0.3) .. (0,0) .. controls (3.31,0.3) and (6.95,1.4) .. (10.93,3.29)   ;
%Straight Lines [id:da9731496520921294] 
\draw  [dash pattern={on 0.84pt off 2.51pt}]  (309.91,127.8) -- (309.91,146.55) ;
%Straight Lines [id:da3650704183908595] 
\draw    (339,101.31) -- (358.47,101.31) ;
\draw [shift={(360.47,101.31)}, rotate = 180] [color={rgb, 255:red, 0; green, 0; blue, 0 }  ][line width=0.75]    (10.93,-3.29) .. controls (6.95,-1.4) and (3.31,-0.3) .. (0,0) .. controls (3.31,0.3) and (6.95,1.4) .. (10.93,3.29)   ;
%Straight Lines [id:da9099325298521992] 
\draw  [dash pattern={on 0.84pt off 2.51pt}]  (384.28,127.62) -- (384.28,146.37) ;
%Straight Lines [id:da24543148806120163] 
\draw    (341.45,171.96) -- (358.47,171.96) ;
\draw [shift={(360.47,171.96)}, rotate = 180] [color={rgb, 255:red, 0; green, 0; blue, 0 }  ][line width=0.75]    (10.93,-3.29) .. controls (6.95,-1.4) and (3.31,-0.3) .. (0,0) .. controls (3.31,0.3) and (6.95,1.4) .. (10.93,3.29)   ;
%Straight Lines [id:da26016373408438453] 
\draw    (411.51,164.74) -- (422.35,152.34) ;
\draw [shift={(423.67,150.83)}, rotate = 131.15] [color={rgb, 255:red, 0; green, 0; blue, 0 }  ][line width=0.75]    (10.93,-3.29) .. controls (6.95,-1.4) and (3.31,-0.3) .. (0,0) .. controls (3.31,0.3) and (6.95,1.4) .. (10.93,3.29)   ;
%Straight Lines [id:da785600150621856] 
\draw    (265.94,141.86) -- (279.79,154.76) ;
\draw [shift={(281.25,156.13)}, rotate = 222.97] [color={rgb, 255:red, 0; green, 0; blue, 0 }  ][line width=0.75]    (10.93,-3.29) .. controls (6.95,-1.4) and (3.31,-0.3) .. (0,0) .. controls (3.31,0.3) and (6.95,1.4) .. (10.93,3.29)   ;
%Straight Lines [id:da05954164486566582] 
\draw    (408.3,104.01) -- (420.47,120.23) ;
\draw [shift={(421.67,121.83)}, rotate = 233.13] [color={rgb, 255:red, 0; green, 0; blue, 0 }  ][line width=0.75]    (10.93,-3.29) .. controls (6.95,-1.4) and (3.31,-0.3) .. (0,0) .. controls (3.31,0.3) and (6.95,1.4) .. (10.93,3.29)   ;
%Straight Lines [id:da34178216073702794] 
\draw    (265.94,130.33) -- (281.66,118.34) ;
\draw [shift={(283.25,117.13)}, rotate = 142.67] [color={rgb, 255:red, 0; green, 0; blue, 0 }  ][line width=0.75]    (10.93,-3.29) .. controls (6.95,-1.4) and (3.31,-0.3) .. (0,0) .. controls (3.31,0.3) and (6.95,1.4) .. (10.93,3.29)   ;
%Straight Lines [id:da37532642126528803] 
\draw    (445,136.81) -- (464.33,136.81) ;
\draw [shift={(466.33,136.81)}, rotate = 180] [color={rgb, 255:red, 0; green, 0; blue, 0 }  ][line width=0.75]    (10.93,-3.29) .. controls (6.95,-1.4) and (3.31,-0.3) .. (0,0) .. controls (3.31,0.3) and (6.95,1.4) .. (10.93,3.29)   ;

% Text Node
\draw (204.15,137) node  [font=\footnotesize]  {$\tilde{x}_{k}^{[ i]}$};
% Text Node
\draw    (241.66,126.15) -- (263.66,126.15) -- (263.66,148.15) -- (241.66,148.15) -- cycle  ;
\draw (252.66,137.15) node  [font=\footnotesize]  {$\Lambda _{i}$};
% Text Node
\draw (310.78,171.94) node  [font=\footnotesize]  {$\begin{bmatrix}
x_{k}^{[ i]}\\
x_{k}^{[ g( q_{i} ,1)]}
\end{bmatrix}$};
% Text Node
\draw (310.95,101.15) node  [font=\footnotesize]  {$\begin{bmatrix}
x_{k}^{[ i]}\\
x_{k}^{[ g( i,1)]}
\end{bmatrix}$};
% Text Node
\draw    (363.28,90.47) -- (405.28,90.47) -- (405.28,112.47) -- (363.28,112.47) -- cycle  ;
\draw (384.28,101.47) node  [font=\footnotesize]  {$\pi _{ig( i,1)}$};
% Text Node
\draw    (363.27,161.08) -- (409.27,161.08) -- (409.27,184.08) -- (363.27,184.08) -- cycle  ;
\draw (386.27,172.58) node  [font=\footnotesize]  {$\pi _{ig( q_{i} ,1)}$};
% Text Node
\draw    (420.58,126.15) -- (443.58,126.15) -- (443.58,148.15) -- (420.58,148.15) -- cycle  ;
\draw (432.08,137.15) node  [font=\footnotesize]  {$\Omega _{i}$};
% Text Node
\draw (491.47,137.76) node  [font=\footnotesize]  {$\Pi _{i}\left(\tilde{x}_{k}^{[ i]}\right)$};
% Text Node
\draw (430.49,93.52) node  [font=\large,color={rgb, 255:red, 74; green, 144; blue, 226 }  ,opacity=1 ]  {$\Pi _{i}$};

\end{tikzpicture}

  \caption{Reformulation of individual MLPs into one larger MLP}
  \label{fig:reformulation}
\end{figure}

To simplify (\ref{eq:controlInput}), we aim to represent it as the mapping of $\tilde{x}^{[i]}_k$ through a single MLP $\Pi_i : \mathbb{R}^{(1+q_i)n_x} \rightarrow \mathbb{R}^{n_u}$. Note that the layer sizes can differ from agent to agent, depending on $|\mathcal{N}_i|$. The mapping from $\tilde{x}^{[i]}_k$ to $\Pi_i(\tilde{x}^{[i]}_k)$ is then
\begin{subequations}
\begin{align}
    z^0_{i,k} &= \tilde{x}^{[i]}_k, \label{eq:NNoverall1}\\ 
    z^{\ell+1}_{i,k} &= \sigma^{\ell}_i\left( W^{\ell}_{i} z^{\ell}_{i,k} + b^{\ell}_{i} \right), \quad \ell = 0,\dots,L-1, \label{eq:NNoverall2}\\ 
    \Pi_{i}\left(\tilde{x}^{[i]}_k\right) &= W^{L}_{i} z^{L}_{i,k} + b^{L}_{i}, \label{eq:NNoverall3}
\end{align}
\end{subequations}
where $z^{\ell}_{i,k} \in \mathbb{R}^{\tilde{n}^i_{\ell}}$ is the $\ell^{\text{th}}$ vector of activation values (note that $\tilde{n}^i_0 = (1+q_i)n_x$), $\sigma^{\ell}_i : \mathbb{R}^{\tilde{n}^i_{\ell+1}} \rightarrow \mathbb{R}^{\tilde{n}^i_{\ell+1}}$ is the $\ell^{\text{th}}$ ReLU activation function, and the weight matrices and bias vectors are given by $W^0_i = \mathfrak{W}^0_i\Lambda_i$, $b^0_i = \mathfrak{b}^0_i$, $W^{\ell}_i = \mathfrak{W}^{\ell}_i $, $b^{\ell}_i = \mathfrak{b}^{\ell}_i$ for $\ell = 1,\dots,L-1$, and $W^L_i = \Omega_i\mathfrak{W}^L_i $, $b^L_i = \Omega_i\mathfrak{b}^L_i$, where
\begin{align*}
    \mathfrak{W}^{\ell}_i &= \mathrm{blkdiag}\left(W^{\ell}_{ig(i,1)},\dots,W^{\ell}_{ig(i,q_i)}\right),\\
    \mathfrak{b}^{\ell}_i &= \rowthree{{b^{\ell}_{ig(i,1)}}^{\top}}{\cdots}{{b^{\ell}_{ig(i,q_i)}}^{\top}}^{\top},
\end{align*}
for $\ell = 0,\dots,L$,
% \begin{align*}
%     W^0_i &= \mathrm{blkdiag}\left(W^0_{ig(i,1)},\dots,W^0_{ig(i,q_i)}\right)\Lambda_i, \\
%     b^0_i &= \rowthree{{b^{0}_{ig(i,1)}}^{\top}}{\cdots}{{b^{0}_{ig(i,q_i)}}^{\top}}^{\top},
% \end{align*}
% and
% \begin{align*}
%     W^{\ell}_i &= \mathrm{blkdiag}\left(W^{\ell}_{ig(i,1)},\dots,W^{\ell}_{ig(i,q_i)}\right),\\
%     b^{\ell}_i &= \rowthree{{b^{\ell}_{ig(i,1)}}^{\top}}{\cdots}{{b^{\ell}_{ig(i,q_i)}}^{\top}}^{\top},
% \end{align*}
% for $\ell = 1,\dots,L-1$, and
% \begin{align*}
%     W^L_i &= \Omega_i\mathrm{blkdiag}\left(W^L_{ig(i,1)},\dots,W^L_{ig(i,q_i)}\right),\\
%     b^L_i &= \Omega_i\rowthree{{b^{L}_{ig(i,1)}}^{\top}}{\cdots}{{b^{L}_{ig(i,q_i)}}^{\top}}^{\top},
% \end{align*}
where $\Lambda_i \in \mathbb{R}^{2q_in_x \times \tilde{n}^i_0}$ and $ \Omega_i \in \mathbb{R}^{n_u \times q_in_u}$ are given by
\begin{equation*}
    \Lambda_i = \begin{bmatrix} I_{n_x} & 0 & 0 & 0 & \cdots & 0 \\
                                0 & I_{n_x} & 0 & 0 & \cdots & 0 \\
                                I_{n_x} & 0 & 0 & 0 & \cdots & 0 \\
                                0 & 0 & I_{n_x} & 0 & \cdots & 0 \\
                                \vdots & \vdots & \vdots & \vdots & \ddots & \vdots \\
                                I_{n_x} & 0 & 0 & 0 & \cdots & 0 \\
                                0 & 0 & 0 & 0 & \cdots & I_{n_x} \\\end{bmatrix},\quad
                                \Omega_i = \colthree{I_{n_u}}{\vdots}{I_{n_u}}^{\top}.
\end{equation*}
Finally, we can write
\begin{equation*}
    \sum_{j \in \mathcal{N}_i} \pi_{ij}\left(\coltwo{x^{[i]}_k}{x^{[j]}_k}\right) \equiv \Pi_i\left(\tilde{x}^{[i]}_k\right),
\end{equation*}
which is illustrated in Figure \ref{fig:reformulation}.

\subsection{Summary of reformulation}

\begin{lemma} \label{lemma:reform}
The dynamics given by (\ref{eq:agentDynamics}) and (\ref{eq:controlInput}) can be written as
\begin{equation}
    x^{[i]}_{k+1} = \tilde{A}_i\tilde{x}^{[i]}_k + B_i\mathrm{sat}_{\mathcal{U}_i}\left[ \Pi_i\left(\tilde{x}^{[i]}_k\right) \right] + w^{[i]}_k, \label{eq:newDynamics}
\end{equation}
where $\tilde{A}_i$ and $\tilde{x}^{[i]}_k$ are defined in (\ref{eq:Atilde}) and (\ref{eq:xtilde}), respectively, and $\Pi_i(\tilde{x}^{[i]}_k)$ is given by (\ref{eq:NNoverall1})--(\ref{eq:NNoverall3}).
\end{lemma}

% \textit{Remark 3:} As suggested in Remark 2, it may be possible that the dynamics may be originally presented in the form in (\ref{eq:newDynamics}), rather than in the form in (\ref{eq:agentDynamics}) and (\ref{eq:controlInput}); in which case, there is no need for a reformulation, and we can proceed with the dynamics in (\ref{eq:newDynamics}).

%%%%%%%%%%%%%%%%%%%%%%%%%%%%%%%%%%%%%%%%%%%%%%%%%%%%%%%%%%%%%%%%%%%%%%%%%%%%%%%%
\section{FORWARD REACHABILITY ANALYSIS} \label{sec:reach}

To overapproximate the forward reachable sets, we extend the reachability method outlined in \cite{002} and \cite{007} for the multi-agent case. The method in \cite{007}, called Reach-SDP, uses quadratic constraints (QCs) and semidefinite programming; the input set, NN and reachable set are abstracted using QCs. This involves forming quadratic inequalities for sets and functions by pre- and post-multiplying a matrix by a `basis vector'. Using `change-of-basis' matrices allows the same basis vector to be used for all inequalities, which allows linear matrix inequalities (LMIs) to be formed for forward reachability analysis of the closed-loop system.

\subsection{Incorporation of control limits into MLP}

We first add two layers to the MLP in (\ref{eq:NNoverall1})--(\ref{eq:NNoverall3}) to account for the control limits \cite{007}. Recall that $\tilde{x}^{[i]}_k$ is the concatenation of the state of agent $i$ and the states of the neighbours of agent $i$ at time $k$. The mapping from $\tilde{x}^{[i]}_k$ to $u^{[i]}_k$ is now defined by (\ref{eq:NNoverall1}), (\ref{eq:NNoverall2}) and 
\begin{subequations}
\begin{align}
    z^{L+1}_{i,k} &=  \sigma^{L}_i\left(W^{L}_{i} z^{L}_{i,k} + b^{L}_{i} - \underline{u}_i \right), \\
    z^{L+2}_{i,k} &= \sigma^{L+1}_i\left(-z^{L+1}_{i,k} + \overline{u}_i - \underline{u}_i \right), \label{eq:extendedMLP4}\\
    u^{[i]}_k &= -z^{L+2}_{i,k} + \overline{u}_i. \label{eq:extendedMLP5}
\end{align}
\end{subequations}
We then define the overall basis vector for the QCs as $\rowtwo{{\mathbf{z}_{i,k}}^{\top}}{1}^{\top}$, where $\mathbf{z}_{i,k}^{\top} = \begin{bmatrix} {z^0_{i,k}}^{\top} & {z^1_{i,k}}^{\top} & \cdots & {z^{L+2}_{i,k}}^{\top} \end{bmatrix}$.

\subsection{Input set}

We describe the input set $\widetilde{\mathcal{X}}^{[i]}_k$, such that $\tilde{x}^{[i]}_k \in \widetilde{\mathcal{X}}^{[i]}_k$, by a hyper-rectangle, i.e.
\begin{equation}
    \widetilde{\mathcal{X}}^{[i]}_k = \left\{ x\in \mathbb{R}^{\tilde{n}^i_0}\ |\ \underline{\tilde{x}}^{[i]}_k \leq x \leq \overline{\tilde{x}}^{[i]}_k \right\}, \label{eq:inputSet}
\end{equation}
where $\underline{\tilde{x}}^{[i]}_k, \overline{\tilde{x}}^{[i]}_k \in \mathbb{R}^{\tilde{n}^i_0}$ are known bounds on the state of the agent and the states of its neighbours at time $k$. Using \cite[Definition 1]{002} and \cite[Proposition 1]{002}, and noting that the first activation value $z^0_{i,k}$ is equal to $\tilde{x}^{[i]}_k$, we can write (\ref{eq:inputSet}) as
\begin{equation}
      \coltwo{z^0_{i,k}}{1}^{\top} P^i_k(\Gamma) \coltwo{z^0_{i,k}}{1} \geq 0,  \label{eq:inputSet2}
\end{equation}
where
\begin{equation*}
    P^i_k(\Gamma) = \mattwo{-2\Gamma}{\Gamma\left(\underline{\tilde{x}}^{[i]}_k+\overline{\tilde{x}}^{[i]}_k\right)}{\left(\underline{\tilde{x}}^{[i]}_k+\overline{\tilde{x}}^{[i]}_k\right)^{\top}\Gamma}{-2{\underline{\tilde{x}}^{[i]}_k}^{\top}\Gamma\overline{\tilde{x}}^{[i]}_k},
\end{equation*}
where $\Gamma \in \mathbb{D}^{\tilde{n}^i_0}$ and $\Gamma \geq 0$. By using a change-of-basis matrix \cite{007}
\begin{equation*}
    E^i_{\mathrm{in}} = \begin{bmatrix} I_{\tilde{n}^i_0} & 0 & \cdots & 0 & 0 \\
                                    0 & 0 & \cdots & 0 & 1
    \end{bmatrix}, 
\end{equation*}
we can write (\ref{eq:inputSet2}) as
\begin{equation}
     \coltwo{\mathbf{z}_{i,k}}{1}^{\top} \Delta^i_k(\Gamma) \coltwo{\mathbf{z}_{i,k}}{1} \geq 0, \label{eq:inputSet3}
\end{equation}
where $\Delta^i_k(\Gamma) = {E^i_{\mathrm{in}}}^{\top} P^i_k(\Gamma) E^i_{\mathrm{in}} $. This is summarised in the following lemma.

\begin{lemma} \label{lemma:QCin}
Consider the state of agent $i$ and the states of the neighbours of agent $i$ at time $k$. If these are are bounded by a hyper-rectangular input set, i.e. $\underline{\tilde{x}}^{[i]}_k \leq \tilde{x}^{[i]}_k \leq \overline{\tilde{x}}^{[i]}_k$, as in (\ref{eq:inputSet}), then (\ref{eq:inputSet3}) holds.
\end{lemma}

\subsection{ReLU activation functions}

First, we define $\hat{z}^{\ell+1}_{i,k} =  W^{\ell}_{i} z^{\ell}_{i,k} + b^{\ell}_{i}$ for $\ell = 0,\dots,L-1$, $\hat{z}^{L+1}_{i,k} = W^{L}_{i} z^{L}_{i,k} + b^{L}_{i} - \underline{u}_i$, and $\hat{z}^{L+2}_{i,k} = -z^{L+1}_{i,k} + \overline{u}_i - \underline{u}_i$. Then, we can write $z^{\ell}_{i,k} = \sigma^{\ell-1}_i(\hat{z}^{\ell}_{i,k})$ for $\ell = 1,\dots,L+2$, and write the concatenation of activation functions as
\begin{equation}
    \boldsymbol{\nu}_{i,k} = \sigma_i\left(\hat{\boldsymbol{\nu}}_{i,k}\right), \label{eq:ReLUoverall}
\end{equation}
where $\boldsymbol{\nu}^{\top}_{i,k} = \rowthree{{z^{1^{\top}}_{i,k}}}{\cdots}{{z^{{L+2}^{\top}}_{i,k}}}$, $\hat{\boldsymbol{\nu}}^{\top}_{i,k} = \rowthree{{\hat{z}^{1^{\top}}_{i,k}}}{\cdots}{{{\hat{z}^{{L+2}^{\top}}_{i,k}}}}$, and $\sigma_i : \mathbb{R}^{n^i_n+2n_u} \rightarrow \mathbb{R}^{n^i_n+2n_u}$ is applied elementwise, where $n^i_n = \sum_{\ell=1}^L\tilde{n}^i_{\ell}$. Using \cite[Definition 2]{002} and \cite[Lemma 3]{002}, we can relax (\ref{eq:ReLUoverall}) as
\begin{equation}
    \colthree{\hat{\boldsymbol{\nu}}_{i,k}}{\boldsymbol{\nu}_{i,k}}{1}^{\top} Q^i(\lambda,\nu,\eta) \colthree{\hat{\boldsymbol{\nu}}_{i,k}}{\boldsymbol{\nu}_{i,k}}{1} \label{eq:ReLUQC} \geq 0,
\end{equation}
where $Q^i(\lambda,\nu,\eta) \in \mathbb{S}^{2(n^i_n+2n_u)+1}$ is defined in \cite[Lemma 3]{002} as $Q$. %Note that $Q^i$ in this paper corresponds to $Q$ in \cite{002}, with $n$ in \cite{002} related by $n = n^i_n+2n_u$.
By using a change-of-base matrix \cite{007}
\begin{equation*}
    E^i_{\mathrm{mid}} = \left[\begin{array}{@{}c|c@{}}
    \begin{matrix}
        W^0_i & 0 & \cdots & 0 & 0 & 0 \\
        0 & W^1_i & \cdots & 0 & 0 & 0 \\
        \vdots & \vdots & \ddots & \vdots & \vdots & \vdots \\
        0 & 0 & \cdots & W^L_i & 0 & 0 \\
        0 & 0 & \cdots & 0 & -I_{n_u} & 0
    \end{matrix} &
  \begin{matrix} 
        {b^0_i} \\ \vdots \\ {b^{L-1}_i} \\ {b^L_i-\underline{u}_i} \\ {\overline{u}_i-\underline{u}_i}
    \end{matrix} \\
\hline
  \begin{matrix} 
        0 & I_{\tilde{n}^i_1} & \cdots & 0 & 0 & 0 \\
        \vdots & \vdots & \ddots & \vdots & \vdots & 0 \\
        0 & 0 & \cdots & I_{\tilde{n}^i_L} & 0 & 0 \\
        0 & 0 & \cdots & 0 & I_{n_u} & 0 \\
        0 & 0 & \cdots & 0 & 0 & I_{n_u} \\
    \end{matrix} & 
  \bigzero \\ \hline \bigzero & 1
\end{array}\right],
\end{equation*}
we can write (\ref{eq:ReLUQC}) as
\begin{equation}
     \coltwo{\mathbf{z}_{i,k}}{1}^{\top} \Theta^i(\lambda,\nu,\eta) \coltwo{\mathbf{z}_{i,k}}{1} \geq 0, \label{eq:ReLUQC2}
\end{equation}
where $\Theta^i(\lambda,\nu,\eta) = {E^i_{\mathrm{mid}}}^{\top} Q^i(\lambda,\nu,\eta) E^i_{\mathrm{mid}} $. This is summarised in the following lemma.

\begin{lemma} \label{lemma:QCmid} Given that the activation values for the extended NN satisfy $z^{\ell}_{i,k} = \sigma^{\ell-1}_i(\hat{z}^{\ell}_{i,k})$ for $\ell = 1,\dots,L+2$, as in (\ref{eq:ReLUoverall}), then (\ref{eq:ReLUQC2}) holds. 
\end{lemma}

\subsection{Reachable set}

We parameterise the overapproximation $\widehat{\mathcal{X}}^{[i]}_{k+1}$ of the reachable set $ {\mathcal{X}}^{[i]}_{k+1}$ using a polytope, i.e. the intersection of $m$ halfspaces
\begin{equation}
    \widehat{\mathcal{X}}^{[i]}_{k+1} = \left\{ x \in \mathbb{R}^{n_x}\ |\ H^{\top}_1x \leq h_1, \dots, H^{\top}_mx \leq h_m \right\}, \label{eq:outputSet1}
\end{equation}
where $H_1,\dots,H_m \in \mathbb{R}^{n_x}$ and $h_1,\dots,h_m \in \mathbb{R}$, which can be written as \cite{007}
\begin{equation*}
    \widehat{\mathcal{X}}^{[i]}_{k+1} = \bigcap_{p=1}^m \left\{ x \in \mathbb{R}^{n_x}\ |\ \coltwo{x}{1}^{\top}S_p(h_p)\coltwo{x}{1} \leq 
    0 \right\},
\end{equation*}
where
\begin{equation*}
    S_p(h_p) = \mattwo{0}{H_p}{H^{\top}_p}{-2h_p},
\end{equation*}
so each halfspace can be equivalently written as
\begin{equation}
    \coltwo{x^{[i]}_{k+1}}{1}^{\top} S_p(h_p) \coltwo{x^{[i]}_{k+1}}{1} \leq 0. \label{eq:outputSet2}
\end{equation}
By using a change-of-basis matrix
\begin{equation*}
    E^i_{\mathrm{out},k} = \begin{bmatrix}
        \tilde{A}_i & 0 & \cdots & 0 & -B_i & B_i\overline{u}_i+w^{[i]}_k \\
        0 & 0 & \cdots & 0 & 0 & 1
    \end{bmatrix},
\end{equation*}
we can write (\ref{eq:outputSet2}) as
\begin{equation}
    \coltwo{\mathbf{z}_{i,k}}{1}^{\top} \Psi^i_k(h_p) \coltwo{\mathbf{z}_{i,k}}{1} \leq 0, \label{eq:outputSet3}
\end{equation}
where $\Psi^i_k(h_p) = {E^i_{\mathrm{out},k}}^{\top}S_p(h_p)E^i_{\mathrm{out},k}$. Note that the structure of the change-of-basis matrix differs to that in \cite{007}, as $\tilde{A}_i$ is not square, to account for the multi-agent dynamics. The result is summarised in the following lemma.

\begin{lemma} \label{lemma:QCout}
% Consider the overapproximation of the reachable set as a polytope, as in (\ref{eq:outputSet1}). Then, (\ref{eq:outputSet3}) holds.
If (\ref{eq:outputSet3}) holds for $p = 1,\dots,m$, then $\widehat{\mathcal{X}}^{[i]}_{k+1}$ is a polytope defined by $H_1,\dots,H_m$ and $h_1,\dots,h_m$, as in (\ref{eq:outputSet1}).
\end{lemma}

\subsection{Reachability algorithm}

By combining the results in Lemmas \ref{lemma:reform}--\ref{lemma:QCout}, we arrive at the following result for the overapproximation of the forward reachable set of agent $i$.

\begin{theorem} \label{theorem:reach}
Consider a discrete-time LTI system of the form in (\ref{eq:agentDynamics}) and (\ref{eq:controlInput}), where the structure of each MLP is given by (\ref{eq:MLP1})--(\ref{eq:MLP3}). At time $k$, let the state of agent $i$ and the states of its neighbours be bounded by a hyper-rectangular input set, as in (\ref{eq:inputSet}). If there exists a solution to
\begin{equation*}\begin{aligned}
    \min_{\Gamma,\lambda,\nu,\eta,h_p} &\quad h_p \\
    \text{subject to} &\quad \Delta^i_k(\Gamma) + \Theta^i(\lambda,\nu,\eta) + \Psi^i_k(h_p) \preceq 0,
\end{aligned}\end{equation*}
for $p = 1,\dots,m$, where $H_1,\dots,H_m$ are specified by the user, then the resulting polytope is the solution to (\ref{eq:originalProblemA})--(\ref{eq:originalProblemC}).
\end{theorem}

\textit{Proof:} Note that we are performing reachability analysis on the system given in (\ref{eq:newDynamics}). However, from Lemma \ref{lemma:reform}, the original dynamics of the form given in (\ref{eq:agentDynamics}) and (\ref{eq:controlInput}) are equivalent to those given in (\ref{eq:newDynamics}), so performing reachability analysis on the system given in (\ref{eq:newDynamics}) is identical to performing reachability analysis on the system given in (\ref{eq:agentDynamics}) and (\ref{eq:controlInput}). The remainder of the proof is similar to that in \cite[Theorem 1]{007}. First, we pre-multiply each of the three terms in the LMI by $\rowtwo{{\mathbf{z}_{i,k}}^{\top}}{1}$ and post-multiply each term by $\rowtwo{{\mathbf{z}_{i,k}}^{\top}}{1}^{\top}$, resulting the scalar inequality
\begin{multline*}
    \coltwo{\mathbf{z}_{i,k}}{1}^{\top} \Delta^i_k(\Gamma)\coltwo{\mathbf{z}_{i,k}}{1} + \coltwo{\mathbf{z}_{i,k}}{1}^{\top}\Theta^i(\lambda,\nu,\eta)\coltwo{\mathbf{z}_{i,k}}{1} \\+ \coltwo{\mathbf{z}_{i,k}}{1}^{\top}\Psi^i_k(h_p)\coltwo{\mathbf{z}_{i,k}}{1} \leq 0.
\end{multline*}
From Lemmas \ref{lemma:QCin} and \ref{lemma:QCmid} and our definitions of the input set and MLP, we know that the first two terms are non-negative, and as the overall expression is non-positive, then the last term is non-positive, i.e.
\begin{equation*}
    \coltwo{\mathbf{z}_{i,k}}{1}^{\top}\Psi^i_k(h_p)\coltwo{\mathbf{z}_{i,k}}{1} \leq 0.
\end{equation*}
Hence, from Lemma \ref{lemma:QCout}, we can conclude that $\widehat{\mathcal{X}}^{[i]}_{k+1}$ is a polytope defined by  $H_1,\dots,H_m$ and $h_1,\dots,h_m$, where $\widehat{\mathcal{X}}^{[i]}_{k+1} \supseteq \mathcal{X}^{[i]}_{k+1}$. \hspace*{\fill} $\blacksquare$

This result is implemented in Algorithm \ref{alg:reachAlgo}, which presents a method for computing hyper-rectangular output sets (a special case of the polytopic sets), given hyper-rectangular input sets. $\mathrm{Reach}( \widetilde{\mathcal{X}}^{[i]}_k)$ represents the solution to the semidefinite program in Theorem 1, given $\widetilde{\mathcal{X}}^{[i]}_k$, and $\delta_{\cdot,\cdot}$ is the Kronecker delta. This algorithm can be used iteratively to find approximate reachable sets for $k+2$, $k+3$, etc. Note also that this approach is parallelisable across agents.

\begin{algorithm} 
\caption{One-step forward reachability analysis with hyper-rectangular constraints}\label{alg:reachAlgo}
\begin{algorithmic}[1]
\Require input sets $\mathcal{X}^{[1]}_k, \dots, \mathcal{X}^{[M]}_k$
% \For{$n = 1,\dots,N$}
\For{$i = 1,\dots,M$}
\State $\widetilde{\mathcal{X}}^{[i]}_k \gets {\mathcal{X}}^{[i]}_k \times {\mathcal{X}}^{[g(i,1)]}_k \times \dots \times{\mathcal{X}}^{[g(i,q_i)]}_k$
\For{$p = 1,\dots,2n_x$}
\If{$p \leq n_x$}
\State $H_p \gets \rowthree{\delta_{1,p}}{\cdots}{\delta_{n_x,p}}^{\top}$
\Else
\State $H_p \gets -\rowthree{\delta_{n_x+1,p}}{\cdots}{\delta_{2n_x,p}}^{\top}$
\EndIf
\State $h^{[i]}_{p,k+1} \gets \mathrm{Reach}\left( \widetilde{\mathcal{X}}^{[i]}_k \right)$
\EndFor
\State $\overline{\hat{x}}^{[i]}_{k+1} \gets \rowthree{h^{[i]}_{1,k+1}}{\cdots}{h^{[i]}_{n_x,k+1}}^{\top}$
\State $\underline{\hat{x}}^{[i]}_{k+1} \gets -\rowthree{h^{[i]}_{n_x+1,k+1}}{\cdots}{h^{[i]}_{2n_x,k+1}}^{\top}$
\State $\widehat{\mathcal{X}}^{[i]}_{k+1} \gets \left\{ x\in \mathbb{R}^{n_x}\ |\ \underline{\hat{x}}^{[i]}_{k+1} \leq x \leq \overline{\hat{x}}^{[i]}_{k+1} \right\}$
\EndFor
\Ensure approximate reachable sets $\widehat{\mathcal{X}}^{[1]}_{k+1}, \dots, \widehat{\mathcal{X}}^{[M]}_{k+1}$

\end{algorithmic}
\end{algorithm}

\section{MODEL UNCERTAINTY}\label{sec:MU}

In this section, we introduce model uncertainty into the dynamics and extend Theorem 1 to account for this. Consider the dynamics in (\ref{eq:agentDynamics}). Instead of assuming direct knowledge of the state and input matrices, we now assume that they lie in the convex hull of a finite number of matrices, i.e.
\begin{equation}
        A_{ii} \in \mathrm{co}\left\{ A_{ii}^1, \dots, A_{ii}^{C_i} \right\},\quad B_{i} \in \mathrm{co}\left\{ B_{i}^1, \dots, B_{i}^{D_i} \right\}, \label{eq:MUcoAB}
\end{equation}
where $C_i, D_i \in \mathbb{Z}^+$, for $i = 1,\dots,M$. Note that $\Psi^i_k$ in (\ref{eq:outputSet3}) depends on $A_{ii}$ and $B_{i}$, so we can write $\Psi^i_k(h_p;A_{ii},B_i)$.

\begin{theorem}\label{theorem:reachMU}
Consider the formulation in Theorem 1, but with the addition of model uncertainty in the state and input matrices described by (\ref{eq:MUcoAB}). If there exists a solution to
\begin{equation*}\begin{aligned}
    \min_{\Gamma,\lambda,\nu,\eta,h_p} &\quad h_p \\
    \text{subject to} &\quad \Delta^i_k(\Gamma) + \Theta^i(\lambda,\nu,\eta) + \Psi^i_k(h_p;A^{c}_{ii},B^{d}_i) \preceq 0, \\
    &\quad \forall\ c \in \left\{1,\dots,C_i\right\},\ d \in \left\{1,\dots,D_i\right\},
\end{aligned}\end{equation*}
for $p = 1,\dots,m$, where $H_1,\dots,H_m$ are specified by the user, then the resulting polytope is the solution to (\ref{eq:originalProblemA})--(\ref{eq:originalProblemC}).
\end{theorem}

% \textit{Proof:} See X. \hspace*{\fill} $\blacksquare$

\textit{Proof:} The proof relies on the fact that $\Psi^i_k(h_p;A^{c}_{ii},B^{d}_i)$ depends affinely on $A^c_{ii}$ and $B^d_i$. The proof has some similarities to that in \cite{132}, but the forms of the matrices and dynamics differ, so we include the proof for completeness. First, note that we can write (\ref{eq:MUcoAB}) as
\begin{equation}
    A_{ii} = \sum_{c=1}^{C_i} \alpha_c A^{c}_{ii}, \quad B_i = \sum_{d=1}^{D_i} \beta_d B^{d}_i, \label{eq:cvxHullAB}
\end{equation}
where $\sum_{c=1}^{C_i} \alpha_c = 1,\ \alpha_c \geq 0\ \forall c\in \{1,\dots,C_i\}$ and $\sum_{d=1}^{D_i} \beta_d = 1,\ \beta_d \geq 0\ \forall d \in \{1,\dots,D_i\}$. From the definitions of $\Psi^i_k$ and $\tilde{A}_i$, we can write
\begin{equation*}
    \Psi^i_k(h_p;A^{c}_{ii},B^{d}_i) = \mattwo{0}{\Phi^i_k(A^{c}_{ii},B^{d}_i)}{{\Phi^i_k(A^{c}_{ii},B^{d}_i)}^{\top}}{\Xi^i_k(h_p;B^{d}_i)},
\end{equation*}
where
\begin{equation*}
    \Phi^i_k(A^{c}_{ii},B^{d}_i) = \begin{bmatrix}
  {\rowfour{A^c_{ii}}{A_{ig(i,1)}}{\cdots}{A_{ig(i,q_i)}}}^{\top}H_p \\ 0 \\ -{B^d_i}^{\top}H_p
  \end{bmatrix},
\end{equation*}
and
\begin{equation*}
    \Xi^i_k(h_p;B^{d}_i) = 2H^{\top}_p\left( B^d_i\overline{u}_i + w^{[i]}_k \right) - 2h_p,
\end{equation*}
so it can be seen that $\Psi^i_k(h_p;A^{c}_{ii},B^{d}_i)$ depends affinely on $A^{c}_{ii}$ and $B^{d}_i$. Hence, multiplying by $\alpha_c$ and $\beta_d$, summing over $c = 1,\dots,C_i$ and $d = 1,\dots,D_i$, and using $\sum_{c=1}^{C_i} \alpha_c A^{c}_{ii} = A_{ii}$, $\sum_{c=1}^{C_i} \alpha_c = 1$, $\sum_{d=1}^{D_i} \beta_d B^{d}_i = B_i$ and $\sum_{d=1}^{D_i} \beta_d = 1$ from (\ref{eq:cvxHullAB}) gives
\begin{equation}
    \sum_{c=1}^{C_i} \sum_{d=1}^{D_i} \alpha_c\beta_d\Psi^i_k(h_p;A^{c}_{ii},B^{d}_i) = \Psi^i_k(h_p;A_{ii},B_i). \label{eq:cvxSumAB} 
\end{equation}
Finally, multiplying the LMIs in the semidefinite program by $\alpha_c$ and $\beta_d$ and summing over $c = 1,\dots,C_i$ and $d = 1,\dots,D_i$, such that
\begin{multline*}
        \sum_{c=1}^{C_i} \sum_{d=1}^{D_i} \alpha_c\beta_d\left[\Delta^i_k(\Gamma) + \Theta^i(\lambda,\nu,\eta)\right] \\ +  \sum_{c=1}^{C_i} \sum_{d=1}^{D_i} \alpha_c\beta_d\Psi^i_k(h_p;A^{c}_{ii},B^{d}_i) \preceq 0,
\end{multline*}
results in
\begin{equation*}
    \Delta^i_k(\Gamma) + \Theta^i(\lambda,\nu,\eta) + \Psi^i_k(h_p;A_{ii},B_i) \preceq 0, 
\end{equation*}
which follows from the fact that $\sum_{c=1}^{C_i} \sum_{d=1}^{D_i} \alpha_c\beta_d = 1$ and from (\ref{eq:cvxSumAB}). \hspace*{\fill} $\blacksquare$

This result is useful, as we only have to solve a finite number of semidefinite programs to find an overapproximation of the forward reachable set for all convex combinations of $A_{ii}^1, \dots, A_{ii}^{C_i}$ and $B_{i}^1, \dots, B_{i}^{D_i}$.

%%%%%%%%%%%%%%%%%%%%%%%%%%%%%%%%%%%%%%%%%%%%%%%%%%%%%%%%%%%%%%%%%%%%%%%%%%%%%%%%
\section{EXPERIMENTS}\label{sec:experiments}

In this section, we use two realistic examples of multi-agent systems to demonstrate our results. We also give a comparison to the approach proposed at the start of Section \ref{sec:reformDyn}, in which the states of each agent are augmented into one overall state and existing reachability methods are applied on the overall system. We also introduce model uncertainty into one of the systems and analyse this case. Simulations were performed on MATLAB, and CVX with MOSEK was used to solve the semidefinite programs. The NNs were trained to approximate distributed MPC schemes with a given horizon; the systems were simulated with MPC, and the resulting input-output data pairs were used to train the NNs.

\subsection{Vehicle platooning}\label{subsec:VP}

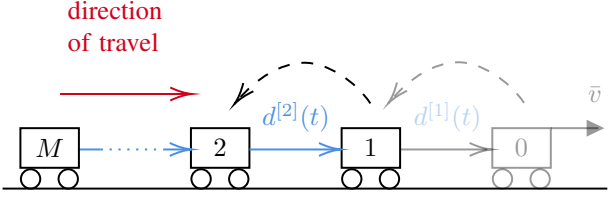
\begin{figure}[h]

\centering

\tikzset{every picture/.style={line width=0.75pt}} %set default line width to 0.75pt        

\begin{tikzpicture}[x=0.75pt,y=0.75pt,yscale=-1,xscale=1]
%uncomment if require: \path (0,300); %set diagram left start at 0, and has height of 300

%Straight Lines [id:da03868761833295098] 
\draw    (153.5,183.75) -- (458.5,183.75) ;
%Shape: Rectangle [id:dp1536527400697134] 
\draw   (324.5,153.25) -- (354,153.25) -- (354,173.25) -- (324.5,173.25) -- cycle ;
%Shape: Circle [id:dp5557770865947609] 
\draw   (324.75,178.88) .. controls (324.75,176.18) and (326.93,174) .. (329.63,174) .. controls (332.32,174) and (334.5,176.18) .. (334.5,178.88) .. controls (334.5,181.57) and (332.32,183.75) .. (329.63,183.75) .. controls (326.93,183.75) and (324.75,181.57) .. (324.75,178.88) -- cycle ;
%Shape: Circle [id:dp677305223625962] 
\draw   (343.75,178.88) .. controls (343.75,176.18) and (345.93,174) .. (348.63,174) .. controls (351.32,174) and (353.5,176.18) .. (353.5,178.88) .. controls (353.5,181.57) and (351.32,183.75) .. (348.63,183.75) .. controls (345.93,183.75) and (343.75,181.57) .. (343.75,178.88) -- cycle ;

%Curve Lines [id:da44286195304447884] 
\draw  [dash pattern={on 4.5pt off 4.5pt}]  (339,140.75) .. controls (331.76,129.77) and (317,119.75) .. (305,120.25) .. controls (293.48,120.73) and (281.96,127.66) .. (271.77,139.5) ;
\draw [shift={(270.5,141)}, rotate = 309.47] [color={rgb, 255:red, 0; green, 0; blue, 0 }  ][line width=0.75]    (10.93,-3.29) .. controls (6.95,-1.4) and (3.31,-0.3) .. (0,0) .. controls (3.31,0.3) and (6.95,1.4) .. (10.93,3.29)   ;
%Shape: Rectangle [id:dp15173753400111445] 
\draw   (248.5,153.25) -- (278,153.25) -- (278,173.25) -- (248.5,173.25) -- cycle ;
%Shape: Circle [id:dp8665159988440738] 
\draw   (248.75,178.88) .. controls (248.75,176.18) and (250.93,174) .. (253.63,174) .. controls (256.32,174) and (258.5,176.18) .. (258.5,178.88) .. controls (258.5,181.57) and (256.32,183.75) .. (253.63,183.75) .. controls (250.93,183.75) and (248.75,181.57) .. (248.75,178.88) -- cycle ;
%Shape: Circle [id:dp2586567072368502] 
\draw   (267.75,178.88) .. controls (267.75,176.18) and (269.93,174) .. (272.63,174) .. controls (275.32,174) and (277.5,176.18) .. (277.5,178.88) .. controls (277.5,181.57) and (275.32,183.75) .. (272.63,183.75) .. controls (269.93,183.75) and (267.75,181.57) .. (267.75,178.88) -- cycle ;

%Straight Lines [id:da30472902003045244] 
\draw [color={rgb, 255:red, 74; green, 144; blue, 226 }  ,draw opacity=1 ]   (192,164) -- (203,164) ;
%Shape: Rectangle [id:dp6169625255486604] 
\draw   (162.5,153.25) -- (192,153.25) -- (192,173.25) -- (162.5,173.25) -- cycle ;
%Shape: Circle [id:dp2696015592904113] 
\draw   (162.75,178.88) .. controls (162.75,176.18) and (164.93,174) .. (167.63,174) .. controls (170.32,174) and (172.5,176.18) .. (172.5,178.88) .. controls (172.5,181.57) and (170.32,183.75) .. (167.63,183.75) .. controls (164.93,183.75) and (162.75,181.57) .. (162.75,178.88) -- cycle ;
%Shape: Circle [id:dp528090521586412] 
\draw   (181.75,178.88) .. controls (181.75,176.18) and (183.93,174) .. (186.63,174) .. controls (189.32,174) and (191.5,176.18) .. (191.5,178.88) .. controls (191.5,181.57) and (189.32,183.75) .. (186.63,183.75) .. controls (183.93,183.75) and (181.75,181.57) .. (181.75,178.88) -- cycle ;

%Straight Lines [id:da40682180694684633] 
\draw [color={rgb, 255:red, 208; green, 2; blue, 27 }  ,draw opacity=1 ]   (182.5,136) -- (247,136) ;
\draw [shift={(249,136)}, rotate = 180] [color={rgb, 255:red, 208; green, 2; blue, 27 }  ,draw opacity=1 ][line width=0.75]    (10.93,-3.29) .. controls (6.95,-1.4) and (3.31,-0.3) .. (0,0) .. controls (3.31,0.3) and (6.95,1.4) .. (10.93,3.29)   ;
%Straight Lines [id:da910316112508585] 
\draw [color={rgb, 255:red, 74; green, 144; blue, 226 }  ,draw opacity=1 ] [dash pattern={on 0.84pt off 2.51pt}]  (203,164) -- (232.5,164) ;
%Straight Lines [id:da9816822383450372] 
\draw [color={rgb, 255:red, 74; green, 144; blue, 226 }  ,draw opacity=1 ]   (234.5,164) -- (246.75,164) ;
\draw [shift={(248.75,164)}, rotate = 180] [color={rgb, 255:red, 74; green, 144; blue, 226 }  ,draw opacity=1 ][line width=0.75]    (10.93,-3.29) .. controls (6.95,-1.4) and (3.31,-0.3) .. (0,0) .. controls (3.31,0.3) and (6.95,1.4) .. (10.93,3.29)   ;
%Straight Lines [id:da6826055929901187] 
\draw [color={rgb, 255:red, 74; green, 144; blue, 226 }  ,draw opacity=1 ]   (278,164) -- (322,164) ;
\draw [shift={(324,164)}, rotate = 180] [color={rgb, 255:red, 74; green, 144; blue, 226 }  ,draw opacity=1 ][line width=0.75]    (10.93,-3.29) .. controls (6.95,-1.4) and (3.31,-0.3) .. (0,0) .. controls (3.31,0.3) and (6.95,1.4) .. (10.93,3.29)   ;
%Shape: Rectangle [id:dp07395838114229258] 
\draw  [color={rgb, 255:red, 0; green, 0; blue, 0 }  ,draw opacity=0.4 ] (400.5,153.25) -- (430,153.25) -- (430,173.25) -- (400.5,173.25) -- cycle ;
%Shape: Circle [id:dp8328929695646241] 
\draw  [color={rgb, 255:red, 0; green, 0; blue, 0 }  ,draw opacity=0.4 ] (400.75,178.88) .. controls (400.75,176.18) and (402.93,174) .. (405.63,174) .. controls (408.32,174) and (410.5,176.18) .. (410.5,178.88) .. controls (410.5,181.57) and (408.32,183.75) .. (405.63,183.75) .. controls (402.93,183.75) and (400.75,181.57) .. (400.75,178.88) -- cycle ;
%Shape: Circle [id:dp38701804204430634] 
\draw  [color={rgb, 255:red, 0; green, 0; blue, 0 }  ,draw opacity=0.4 ] (419.75,178.88) .. controls (419.75,176.18) and (421.93,174) .. (424.63,174) .. controls (427.32,174) and (429.5,176.18) .. (429.5,178.88) .. controls (429.5,181.57) and (427.32,183.75) .. (424.63,183.75) .. controls (421.93,183.75) and (419.75,181.57) .. (419.75,178.88) -- cycle ;

%Straight Lines [id:da9331209723101972] 
\draw [color={rgb, 255:red, 0; green, 0; blue, 0 }  ,draw opacity=0.4 ]   (354,164) -- (398,164) ;
\draw [shift={(400,164)}, rotate = 180] [color={rgb, 255:red, 0; green, 0; blue, 0 }  ,draw opacity=0.4 ][line width=0.75]    (10.93,-3.29) .. controls (6.95,-1.4) and (3.31,-0.3) .. (0,0) .. controls (3.31,0.3) and (6.95,1.4) .. (10.93,3.29)   ;
%Straight Lines [id:da716593962987095] 
\draw [color={rgb, 255:red, 0; green, 0; blue, 0 }  ,draw opacity=0.4 ]   (430,153.25) -- (454,153.25) ;
\draw [shift={(457,153.25)}, rotate = 180] [fill={rgb, 255:red, 0; green, 0; blue, 0 }  ,fill opacity=0.4 ][line width=0.08]  [draw opacity=0] (8.93,-4.29) -- (0,0) -- (8.93,4.29) -- cycle    ;
%Curve Lines [id:da2771384556600105] 
\draw [color={rgb, 255:red, 0; green, 0; blue, 0 }  ,draw opacity=0.4 ] [dash pattern={on 4.5pt off 4.5pt}]  (416.5,140.5) .. controls (409.26,129.52) and (394.5,119.5) .. (382.5,120) .. controls (370.98,120.48) and (359.46,127.41) .. (349.27,139.25) ;
\draw [shift={(348,140.75)}, rotate = 309.47] [color={rgb, 255:red, 0; green, 0; blue, 0 }  ,draw opacity=0.4 ][line width=0.75]    (10.93,-3.29) .. controls (6.95,-1.4) and (3.31,-0.3) .. (0,0) .. controls (3.31,0.3) and (6.95,1.4) .. (10.93,3.29)   ;

% Text Node
\draw (339.25,163.25) node    {$1$};
% Text Node
\draw (177.25,163.25) node    {$M$};
% Text Node
\draw (185,87.5) node [anchor=north west][inner sep=0.75pt]  [color={rgb, 255:red, 208; green, 2; blue, 27 }  ,opacity=1 ] [align=left] {direction\\of travel};
% Text Node
\draw (301.53,147.5) node  [color={rgb, 255:red, 74; green, 144; blue, 226 }  ,opacity=1 ]  {$d^{[ 2]}( t)$};
% Text Node
\draw (415.25,163.25) node  [color={rgb, 255:red, 0; green, 0; blue, 0 }  ,opacity=0.4 ]  {$0$};
% Text Node
\draw (452.25,136.25) node  [color={rgb, 255:red, 0; green, 0; blue, 0 }  ,opacity=0.4 ]  {$\bar{v}$};
% Text Node
\draw (378.03,147.5) node  [color={rgb, 255:red, 74; green, 144; blue, 226 }  ,opacity=0.4 ]  {$d^{[ 1]}( t)$};
% Text Node
\draw (263.25,163.25) node    {$2$};

\end{tikzpicture}

    \caption{Platoon of $M$ vehicles, where each vehicle $i$ only receives information from vehicle $i-1$; the `virtual' vehicle is shown with increased opacity}
    \label{fig:platoon}
\end{figure}

\begin{figure}[h]
    \centering
    \includegraphics[width=0.45\textwidth]{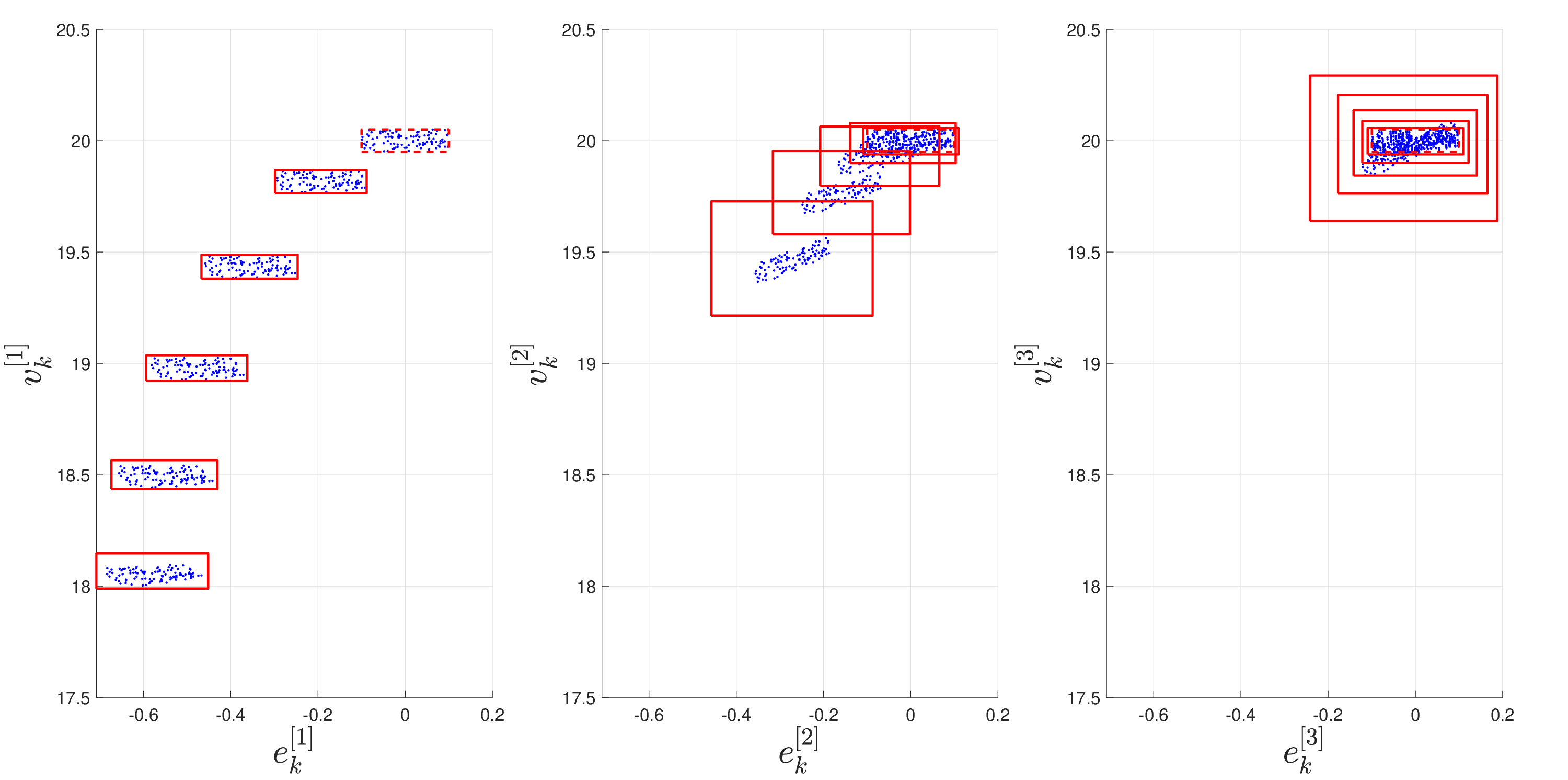}
    \caption{Plots of the reachable sets in red (solid) for distance error and velocity, and simulated trajectories (blue) for the vehicle platooning example; the initial set is shown in red (dashed) -- only agents $1$ to $3$ are shown}
    \label{fig:VPresults}
\end{figure}

In the first example, we consider the example of control of a platoon of vehicles. There are several forms of this problem, and control of a vehicular platoon has a number of benefits, including improved safety, higher road capacity, lower emissions, and/or reduced congestion \cite{069,087,020,025}. In this example, we consider the adaptive cruise control (ACC) problem, in which each vehicle aims to maintain a fixed distance from the vehicle in front, whilst travelling at a given velocity. The continuous-time longitudinal dynamics of each vehicle are given by \cite{024}
\begin{equation*}
    \dot{x}^{[i]}(t) = \bar{A}_{ii}x^{[i]}(t) + \bar{A}_{ii-1}x^{[i-1]}(t) + \bar{B}_{i}u^{[i]}(t),
\end{equation*}
(note that $\mathcal{N}_i = \{i-1\}$) where the state vector is
\begin{equation*}
    x^{[i]}(t) = \rowthree{e^{[i]}(t)}{v^{[i]}(t)}{a^{[i]}(t)}^{\top},
\end{equation*}
where $e^{[i]}(t)$ is the distance error between vehicle $i$ and vehicle $i-1$ (i.e. if the desired distance between vehicle $i$ and the vehicle in front, $i-1$, is $\bar{d}^{[i]}$ and the actual distance is $d^{[i]}(t)$, then $e^{[i]}(t) = d^{[i]}(t)-\bar{d}_i$), $v^{[i]}(t)$ is the velocity of the $i^{\text{th}}$ vehicle, $a^{[i]}(t)$ is the acceleration of the $i^{\text{th}}$ vehicle, and
\begin{equation*}\begin{aligned} 
    \bar{A}_{ii} &= \matthree{0}{-1}{0}{0}{0}{1}{0}{0}{-\frac{1}{\tau}},\quad
    \bar{A}_{ii-1} = \matthree{0}{1}{0}{0}{0}{0}{0}{0}{0},\\
    \bar{B}_i &= \rowthree{0}{0}{\frac{1}{\tau}}^{\top}, 
\end{aligned} \end{equation*}
where $\tau$ is the engine time constant \cite{146} and $u^{[i]}(t)$ is the $i^{\text{th}}$ acceleration input.
Note that the lead vehicle ($i=1$) has no physical neighbour, but this can be resolved by imagining a virtual vehicle \cite{024} with state $x^{[0]} = \rowthree{0}{\bar{v}}{0}^{\top}$, where $\bar{v}$ is the reference velocity (to be maintained by the platoon). This is shown in Figure \ref{fig:platoon}.
The dynamics are discretised assuming zero-order hold (ZOH) with a sample period $T = 0.1\ \mathrm{s}$, treating $x^{[i-1]}(t)$ and $u^{[i]}(t)$ as exogenous inputs, such that the discrete-time dynamics are in the form in (\ref{eq:agentDynamics}), where $w^{[i]}_k = 0\ \forall i \in \mathcal{I}$. The NNs have $2$ hidden layers, both with $15$ neurons. The $i^{\text{th}}$ control input is
\begin{equation*}
    u^{[i]}_k = \mathrm{sat}_{\mathcal{U}_i}\left[\pi_{ii-1  }\left( \colthree{e^{[i]}_k}{v^{[i-1]}_k-v^{[i]}_k}{a^{[i-1]}_k-a^{[i]}_k} \right)\right],
\end{equation*}
where $e^{[i]}_k \equiv e^{[i]}(kT)$, $v^{[i]}_k \equiv v^{[i]}(kT)$ and $a^{[i]}_k \equiv a^{[i]}(kT)$. The forward reachable sets were computed for five time steps, $M = 9$ agents, initial conditions given by $\mathcal{X}^{[i]}_0 = \{x \in \mathbb{R}^3\ |\ \underline{x} \leq x \leq \overline{x} \}\  \forall i \in \mathcal{I}$, where $\underline{x}^{\top} = \rowthree{-0.1}{19.95}{-0.01}$ and $\overline{x}^{\top} = \rowthree{0.1}{20.05}{0.01}$, and controller limits given by $\overline{u}_i = -\underline{u}_i = 5\ \forall i \in \mathcal{I}$. A step change of $-2$ was applied to the reference velocity at $k = 0$ (from $\bar{v} = 20$ to $\bar{v} = 18$). The results are shown in Figure \ref{fig:VPresults} for the first three vehicles.
Note that the step change in $\bar{v}$ takes some time to propagate down the platoon, hence the difference in range between the plots. A comparison between the computation time for this approach (Reach-SDP-MA) and the existing method (Reach-SDP) is shown in Table \ref{tab:comparison} for different values of $M$. 

\begin{table}[h]\caption{Comparison of methods (times in s)} 
\begin{center}
\begin{tabular}{|c||c|c|c|c|c|}
\hline
$M$ & $1$ & $2$ & $3$ & $4$ & $5$ \\
\hline
Reach-SDP-MA & $3.85$ & $7.04$ & $11.60$ & $14.17$ & $20.82$ \\
\hline
Reach-SDP \cite{007} & $4.01$ & $75.39$ & $562.94$ & $3636.62$ & $41025.05$ \\
\hline
\end{tabular}
\end{center}
\label{tab:comparison}
\end{table}

\begin{figure}[h]
    \centering
    \includegraphics[width=0.45\textwidth]{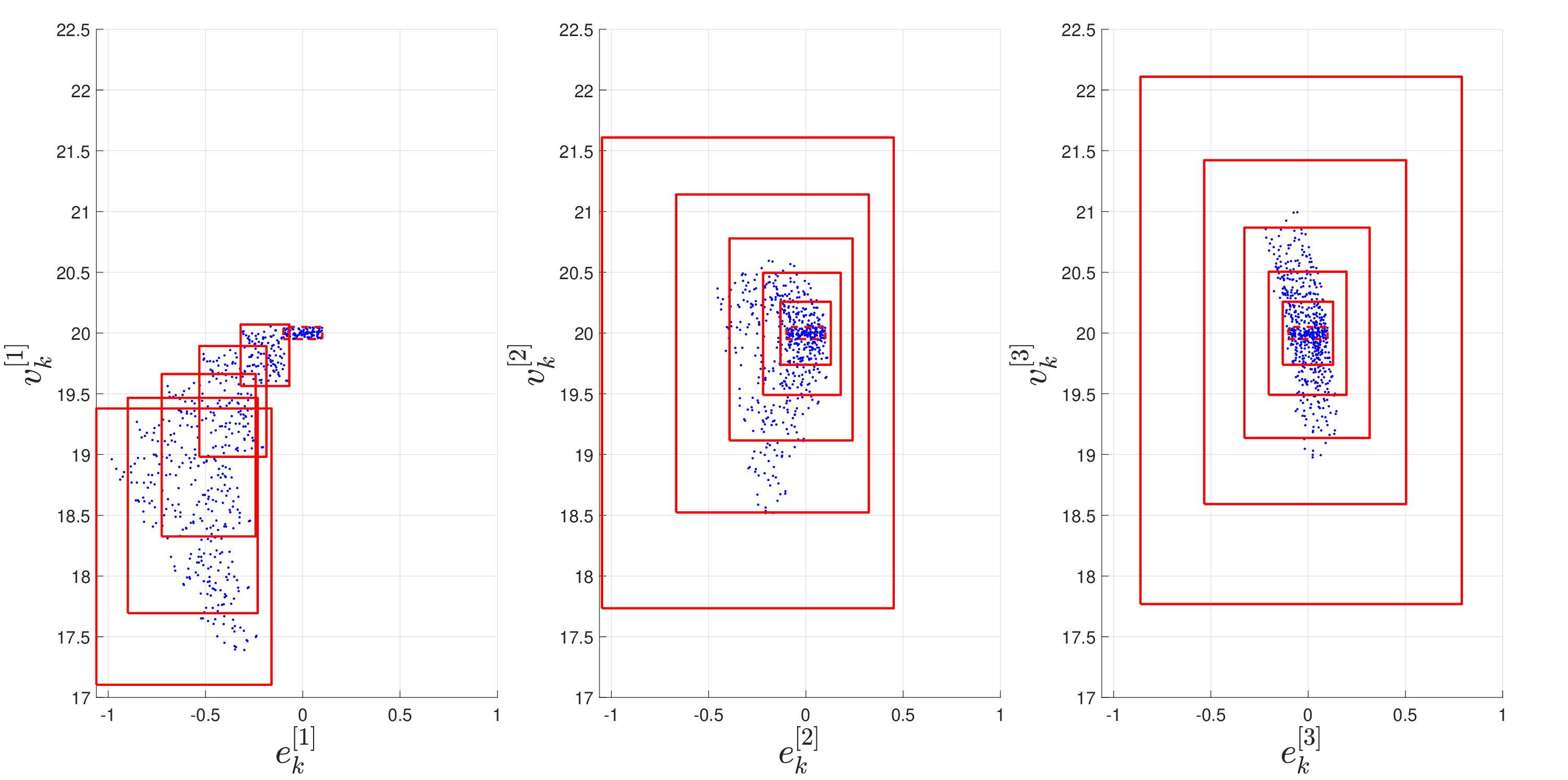}
    \caption{Plots of the reachable sets in red (solid) for distance error and velocity, and simulated trajectories (blue) for the vehicle platooning example with model uncertainty; the initial set is shown in red (dashed) -- only agents $1$ to $3$ are shown}
    \label{fig:VPMUresults}
\end{figure}

We then extend the vehicle platooning example to account for model uncertainty. Consider the case in which $A_{ii} \in \mathrm{co}\left\{ (1-\delta) A^0, (1+\delta) A^0 \right\}$ and $B_i \in \mathrm{co}\left\{ (1-\delta) B^0, (1+\delta) B^0 \right\}$,
% \begin{align*}
%     A_{ii} &\in \mathrm{co}\left\{ (1-\delta) A^0, (1+\delta) A^0 \right\}, \\
%     B_i &\in \mathrm{co}\left\{ (1-\delta) B^0, (1+\delta) B^0 \right\},
% \end{align*}
where $A^0 \in \mathbb{R}^{3\times 3}$ and $B^0 \in \mathbb{R}^{3}$ are the nominal values of $A_{ii}$ and $B_{i}$, respectively, and $\delta = 0.01$.
% This means that we want to find the forward reachable sets for the case in which
% \begin{align*}
%     A_{ii} &= \lambda_1(1-\delta) A^0 + (1-\lambda_1)\delta A^0, \\
%     B_i &= \lambda_2(1-\delta) B^0 +  (1-\lambda_2)\delta B^0,
% \end{align*}
% for all $\lambda_1,\lambda_2 \in [0,1]$.
The results are shown in Figure \ref{fig:VPMUresults} for the first three vehicles.
Because of the uncertainty, the size of the exact reachable sets increases, so the overapproximations are larger (compared to Figure \ref{fig:VPresults}).

\subsection{Power network system}

\begin{figure}[h]
    \centering
    \tikzset{every picture/.style={line width=0.75pt}} %set default line width to 0.75pt        
    \begin{tikzpicture}[x=0.75pt,y=0.75pt,yscale=-1,xscale=1]
    %uncomment if require: \path (0,300); %set diagram left start at 0, and has height of 300
    
    %Shape: Rectangle [id:dp6760039808456355] 
    \draw   (121.25,93.52) -- (161.72,93.52) -- (161.72,116.65) -- (121.25,116.65) -- cycle ;
    %Shape: Rectangle [id:dp08978269961625496] 
    \draw   (121.25,174.1) -- (161.72,174.1) -- (161.72,197.23) -- (121.25,197.23) -- cycle ;
    %Shape: Rectangle [id:dp784518057078617] 
    \draw   (242.3,93.52) -- (282.76,93.52) -- (282.76,116.65) -- (242.3,116.65) -- cycle ;
    %Shape: Rectangle [id:dp826994425428583] 
    \draw   (242.3,174.1) -- (282.76,174.1) -- (282.76,197.23) -- (242.3,197.23) -- cycle ;
    %Straight Lines [id:da18606808258096708] 
    \draw    (262.53,119.65) -- (262.53,171.1) ;
    \draw [shift={(262.53,174.1)}, rotate = 270] [fill={rgb, 255:red, 0; green, 0; blue, 0 }  ][line width=0.08]  [draw opacity=0] (8.93,-4.29) -- (0,0) -- (8.93,4.29) -- cycle    ;
    \draw [shift={(262.53,116.65)}, rotate = 90] [fill={rgb, 255:red, 0; green, 0; blue, 0 }  ][line width=0.08]  [draw opacity=0] (8.93,-4.29) -- (0,0) -- (8.93,4.29) -- cycle    ;
    %Straight Lines [id:da6568996106455873] 
    \draw    (141.48,119.65) -- (141.48,171.1) ;
    \draw [shift={(141.48,174.1)}, rotate = 270] [fill={rgb, 255:red, 0; green, 0; blue, 0 }  ][line width=0.08]  [draw opacity=0] (8.93,-4.29) -- (0,0) -- (8.93,4.29) -- cycle    ;
    \draw [shift={(141.48,116.65)}, rotate = 90] [fill={rgb, 255:red, 0; green, 0; blue, 0 }  ][line width=0.08]  [draw opacity=0] (8.93,-4.29) -- (0,0) -- (8.93,4.29) -- cycle    ;
    %Straight Lines [id:da18256648313459656] 
    \draw    (164.72,111.47) -- (239.3,111.47) ;
    \draw [shift={(242.3,111.47)}, rotate = 180] [fill={rgb, 255:red, 0; green, 0; blue, 0 }  ][line width=0.08]  [draw opacity=0] (8.93,-4.29) -- (0,0) -- (8.93,4.29) -- cycle    ;
    \draw [shift={(161.72,111.47)}, rotate = 0] [fill={rgb, 255:red, 0; green, 0; blue, 0 }  ][line width=0.08]  [draw opacity=0] (8.93,-4.29) -- (0,0) -- (8.93,4.29) -- cycle    ;
    %Straight Lines [id:da050983374755131594] 
    \draw    (161.72,101.09) -- (179.95,101.09) ;
    \draw [shift={(181.95,101.09)}, rotate = 180] [color={rgb, 255:red, 0; green, 0; blue, 0 }  ][line width=0.75]    (10.93,-3.29) .. controls (6.95,-1.4) and (3.31,-0.3) .. (0,0) .. controls (3.31,0.3) and (6.95,1.4) .. (10.93,3.29)   ;
    %Straight Lines [id:da17993665376314638] 
    \draw    (161.72,184.75) -- (179.95,184.75) ;
    \draw [shift={(181.95,184.75)}, rotate = 180] [color={rgb, 255:red, 0; green, 0; blue, 0 }  ][line width=0.75]    (10.93,-3.29) .. controls (6.95,-1.4) and (3.31,-0.3) .. (0,0) .. controls (3.31,0.3) and (6.95,1.4) .. (10.93,3.29)   ;
    %Straight Lines [id:da10774650822029463] 
    \draw    (282.76,105.09) -- (301,105.09) ;
    \draw [shift={(303,105.09)}, rotate = 180] [color={rgb, 255:red, 0; green, 0; blue, 0 }  ][line width=0.75]    (10.93,-3.29) .. controls (6.95,-1.4) and (3.31,-0.3) .. (0,0) .. controls (3.31,0.3) and (6.95,1.4) .. (10.93,3.29)   ;
    %Straight Lines [id:da4891186099046665] 
    \draw    (282.76,184.75) -- (301,184.75) ;
    \draw [shift={(303,184.75)}, rotate = 180] [color={rgb, 255:red, 0; green, 0; blue, 0 }  ][line width=0.75]    (10.93,-3.29) .. controls (6.95,-1.4) and (3.31,-0.3) .. (0,0) .. controls (3.31,0.3) and (6.95,1.4) .. (10.93,3.29)   ;
    %Straight Lines [id:da02078192119804645] 
    \draw    (141.48,75.6) -- (141.48,91.52) ;
    \draw [shift={(141.48,93.52)}, rotate = 270] [color={rgb, 255:red, 0; green, 0; blue, 0 }  ][line width=0.75]    (10.93,-3.29) .. controls (6.95,-1.4) and (3.31,-0.3) .. (0,0) .. controls (3.31,0.3) and (6.95,1.4) .. (10.93,3.29)   ;
    %Straight Lines [id:da02080958951578249] 
    \draw    (262.53,75.6) -- (262.53,91.52) ;
    \draw [shift={(262.53,93.52)}, rotate = 270] [color={rgb, 255:red, 0; green, 0; blue, 0 }  ][line width=0.75]    (10.93,-3.29) .. controls (6.95,-1.4) and (3.31,-0.3) .. (0,0) .. controls (3.31,0.3) and (6.95,1.4) .. (10.93,3.29)   ;
    %Straight Lines [id:da1988148926315627] 
    \draw    (114.86,162.92) -- (127.15,172.85) ;
    \draw [shift={(128.7,174.1)}, rotate = 218.93] [color={rgb, 255:red, 0; green, 0; blue, 0 }  ][line width=0.75]    (10.93,-3.29) .. controls (6.95,-1.4) and (3.31,-0.3) .. (0,0) .. controls (3.31,0.3) and (6.95,1.4) .. (10.93,3.29)   ;
    %Straight Lines [id:da4387113191064014] 
    \draw    (236.62,162.92) -- (248.9,172.85) ;
    \draw [shift={(250.46,174.1)}, rotate = 218.93] [color={rgb, 255:red, 0; green, 0; blue, 0 }  ][line width=0.75]    (10.93,-3.29) .. controls (6.95,-1.4) and (3.31,-0.3) .. (0,0) .. controls (3.31,0.3) and (6.95,1.4) .. (10.93,3.29)   ;
    %Straight Lines [id:da11508564644562891] 
    \draw    (101.02,184.89) -- (119.25,184.89) ;
    \draw [shift={(121.25,184.89)}, rotate = 180] [color={rgb, 255:red, 0; green, 0; blue, 0 }  ][line width=0.75]    (10.93,-3.29) .. controls (6.95,-1.4) and (3.31,-0.3) .. (0,0) .. controls (3.31,0.3) and (6.95,1.4) .. (10.93,3.29)   ;
    %Straight Lines [id:da17580857883307144] 
    \draw    (226.68,195.58) -- (240.64,186.02) ;
    \draw [shift={(242.3,184.89)}, rotate = 145.62] [color={rgb, 255:red, 0; green, 0; blue, 0 }  ][line width=0.75]    (10.93,-3.29) .. controls (6.95,-1.4) and (3.31,-0.3) .. (0,0) .. controls (3.31,0.3) and (6.95,1.4) .. (10.93,3.29)   ;
    %Straight Lines [id:da9234045576662289] 
    \draw    (233.42,82.34) -- (245.71,92.27) ;
    \draw [shift={(247.26,93.52)}, rotate = 218.93] [color={rgb, 255:red, 0; green, 0; blue, 0 }  ][line width=0.75]    (10.93,-3.29) .. controls (6.95,-1.4) and (3.31,-0.3) .. (0,0) .. controls (3.31,0.3) and (6.95,1.4) .. (10.93,3.29)   ;
    %Straight Lines [id:da2378929841854236] 
    \draw    (101.02,105.09) -- (119.25,105.09) ;
    \draw [shift={(121.25,105.09)}, rotate = 180] [color={rgb, 255:red, 0; green, 0; blue, 0 }  ][line width=0.75]    (10.93,-3.29) .. controls (6.95,-1.4) and (3.31,-0.3) .. (0,0) .. controls (3.31,0.3) and (6.95,1.4) .. (10.93,3.29)   ;
    
    % Text Node
    \draw (141.48,185.66) node    {$1$};
    % Text Node
    \draw (141.48,105.09) node    {$2$};
    % Text Node
    \draw (262.53,105.09) node    {$3$};
    % Text Node
    \draw (262.53,185.66) node    {$4$};
    % Text Node
    \draw (126.01,142.65) node    {$\textcolor[rgb]{0.29,0.56,0.89}{P}\textcolor[rgb]{0.29,0.56,0.89}{_{12}}$};
    % Text Node
    \draw (277.74,146.91) node    {$\textcolor[rgb]{0.29,0.56,0.89}{P}\textcolor[rgb]{0.29,0.56,0.89}{_{34}}$};
    % Text Node
    \draw (201.4,124.19) node    {$\textcolor[rgb]{0.29,0.56,0.89}{P}\textcolor[rgb]{0.29,0.56,0.89}{_{23}}$};
    % Text Node
    \draw (203.32,97.13) node    {$\Delta \omega _{k}^{[ 2]}$};
    % Text Node
    \draw (203.86,184.23) node    {$\Delta \omega _{k}^{[ 1]}$};
    % Text Node
    \draw (325.62,104.42) node  [rotate=-359.94]  {$\Delta \omega _{k}^{[ 3]}$};
    % Text Node
    \draw (327.35,184.49) node  [rotate=-359.94]  {$\Delta \omega _{k}^{[ 4]}$};
    % Text Node
    \draw (141.48,65.11) node  [color={rgb, 255:red, 208; green, 2; blue, 27 }  ,opacity=1 ]  {$\Delta P_{L,k}^{[ 2]}$};
    % Text Node
    \draw (262.53,65.11) node  [color={rgb, 255:red, 208; green, 2; blue, 27 }  ,opacity=1 ]  {$\Delta P_{L,k}^{[ 3]}$};
    % Text Node
    \draw (95.99,159.58) node  [color={rgb, 255:red, 208; green, 2; blue, 27 }  ,opacity=1 ]  {$\Delta P_{L}^{[ 1]}$};
    % Text Node
    \draw (221.49,154.41) node  [color={rgb, 255:red, 208; green, 2; blue, 27 }  ,opacity=1 ]  {$\Delta P_{L}^{[ 4]}$};
    % Text Node
    \draw (86.25,184.89) node  [color={rgb, 255:red, 0; green, 0; blue, 0 }  ,opacity=1 ]  {$u_{k}^{[ 1]}$};
    % Text Node
    \draw (221.57,206.82) node  [color={rgb, 255:red, 0; green, 0; blue, 0 }  ,opacity=1 ]  {$u_{k}^{[ 4]}$};
    % Text Node
    \draw (224.92,72.76) node  [color={rgb, 255:red, 0; green, 0; blue, 0 }  ,opacity=1 ]  {$u_{k}^{[ 3]}$};
    % Text Node
    \draw (87.45,102.69) node  [color={rgb, 255:red, 0; green, 0; blue, 0 }  ,opacity=1 ]  {$u_{k}^{[ 2]}$};

    \end{tikzpicture}
    \caption{Power network system with $4$ areas (adapted from \cite{147})}
    \label{fig:PNS}
\end{figure}
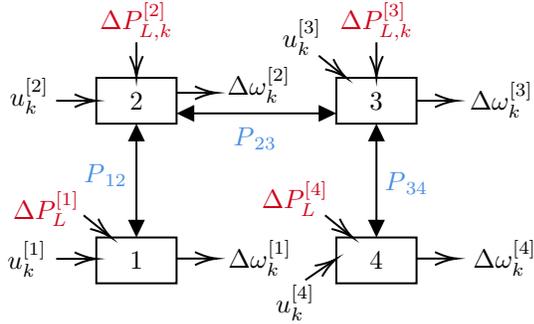

\begin{figure}[h]
    \centering
    \includegraphics[width=0.45\textwidth]{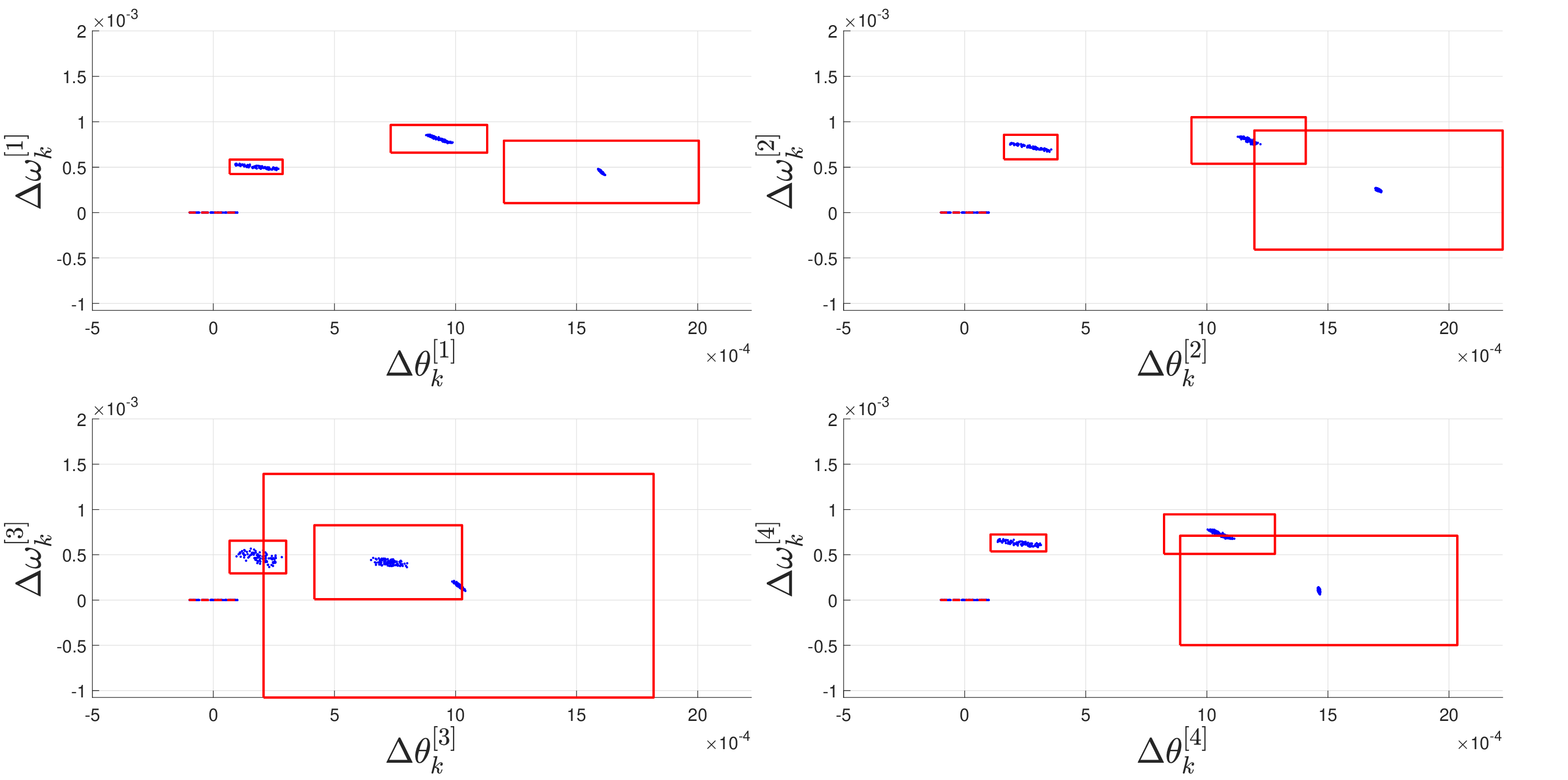}
    \caption{Plots of the reachable sets in red (solid) for angle and frequency deviation and simulated trajectories (blue) for the power network example; the initial set is shown in red (dashed); here, $\Delta\theta^{[i]}_k \equiv \Delta\theta^{[i]}(kT)\ \forall i \in \mathcal{I}$ and $\Delta\omega^{[i]}_k \equiv \Delta\omega^{[i]}(kT)\ \forall i \in \mathcal{I}$ }
    \label{fig:PNresults}
\end{figure}

For the second example, we consider automatic generation control (AGC) of a power network system. Unlike the previous example, the dynamics are not identical across agents, and some agents have more than one neighbour. There is also an exogenous input term. This system consists of $M$ generation areas, and the aim is to reduce the frequency deviation in each area, in spite of load changes. Common approaches to this control problem include decentralised and distributed MPC schemes \cite{147,126,074}. The continuous-time dynamics of each area are given by \cite{147,saadat1999power}
\begin{equation*}
    \dot{x}^{[i]}(t) = \bar{A}_{ii}x^{[i]}(t) + \sum_{j\in \mathcal{N}_i} \bar{A}_{ij}x^{[j]}(t) + \bar{B}_{i}u^{[i]}(t) + \bar{L}_i\Delta P^{[i]}_{L}(t),
\end{equation*}
% \begin{dmath*}
%     \dot{x}^{[i]}(t) = \bar{A}_{ii}x^{[i]}(t) + \sum_{j\in \mathcal{N}_i} \bar{A}_{ij}x^{[j]}(t)\\ + \bar{B}_{i}u^{[i]}(t) + \bar{L}_i\Delta P^{[i]}_{L}(t),
% \end{dmath*}
where the state vector for area $i$ is
\begin{equation*}
    x^{[i]}(t) = \rowfour{\Delta\theta^{[i]}(t)}{\Delta\omega^{[i]}(t)}{\Delta P_{m}^{[i]}(t)}{\Delta P_{v}^{[i]}(t)}^{\top},
\end{equation*}
where $\Delta\theta^{[i]}(t)$, $\Delta\omega^{[i]}(t)$, $\Delta P_{m}^{[i]}(t)$ and $\Delta P_{v}^{[i]}(t)$ are the deviations in rotor angle, frequency, mechanical power and steam valve position, respectively, from the nominal values \cite{096}, $u^{[i]}(t)$ is the reference power, $\Delta P^{[i]}_{L}(t)$ is the local power load, and
\begin{equation*}\begin{aligned} 
    \bar{A}_{ii} &= \begin{bmatrix}
        0 & 1 & 0 & 0 \\ -\frac{\sum_{j\in \mathcal{N}_i}P_{ij}}{2H_i} & -\frac{D_i}{2H_i} & \frac{1}{2H_i} & 0 \\ 0 & 0 & -\frac{1}{T_{t_i}} & \frac{1}{T_{t_i}} \\ 0 & -\frac{1}{R_iT_{g_i}} & 0 & -\frac{1}{T_{g_i}} 
    \end{bmatrix},\\
    \bar{A}_{ij} &= \begin{bmatrix} 
        0 & 0 & 0 & 0 \\ \frac{P_{ij}}{2H_i} & 0 & 0 & 0 \\ 0 & 0 & 0 & 0 \\ 0 & 0 & 0 & 0
    \end{bmatrix},\quad \begin{aligned}
    \bar{B}_i &=  \rowfour{0}{0}{0}{T_{g_i}}^{\top},\\
    \bar{L}_i &= \rowfour{0}{-\frac{1}{2H_i}}{0}{0}^{\top},     \end{aligned}
\end{aligned}  \end{equation*}
where $P_{ij}$, $H_i$, $D_i$, $T_{t_i}$, $R_i$ and $T_{g_i}$ are defined in \cite{147}.
In this example, we consider $M = 4$ generation areas (Scenario 1 in \cite{096}), where $\mathcal{N}_1 = \{2\}$, $\mathcal{N}_2 = \{1,3\}$, $\mathcal{N}_3 = \{2,4\}$ and $\mathcal{N}_4 = \{3\}$, as shown in Figure \ref{fig:PNS}.
The dynamics are discretised assuming ZOH with a sample period $T = 1\ \mathrm{s}$, treating $x^{[j]}(t)\ \forall j \in \mathcal{N}_i$, $u^{[i]}(t)$ and $\Delta P^{[i]}_{L}(t)$ as exogenous inputs, such that the discrete-time dynamics are in the form in (\ref{eq:agentDynamics}). The NNs have $2$ hidden layers, both with $10$ neurons. The $i^{\text{th}}$ control input is
\begin{equation*}
    u^{[i]}_k = \mathrm{sat}_{\mathcal{U}_i}\left[ \sum_{j \in \mathcal{N}_i} \pi_{ij}\left(\coltwo{x^{[i]}_k}{x^{[j]}_k}-\coltwo{x^{[i]}_{\mathrm{ref},k}}{x^{[j]}_{\mathrm{ref},k}}\right) \right],
\end{equation*}
where $x^{[i]}_{\mathrm{ref},k}$ and $x^{[j]}_{\mathrm{ref},k}$ are the state and neighbour reference values, respectively. These are incorporated into the reachability analysis by modifying the first weight matrices and bias vectors. The forward reachable sets were computed for three time steps with initial conditions given by $\mathcal{X}^{[i]}_0 = \{x \in \mathbb{R}^4\ |\ \underline{x} \leq x \leq \overline{x} \}\  \forall i \in \mathcal{I}$, where $\overline{x}^{\top} = -\underline{x}^{\top} = \rowfour{10^{-4}}{10^{-7}}{10^{-3}}{10^{-3}}$, and controller limits given in \cite{096}. A step change of $-0.15$ was applied to $\Delta P_{L,k}^{[i]} \equiv \Delta P_{L}^{[i]}(kT)\ \forall i \in \mathcal{I}$, and the results are shown in Figure \ref{fig:PNresults}.

%%%%%%%%%%%%%%%%%%%%%%%%%%%%%%%%%%%%%%%%%%%%%%%%%%%%%%%%%%%%%%%%%%%%%%%%%%%%%%%%
\section{CONCLUSION AND FUTURE WORK}

In this paper, we presented a scalable method to overapproximate the forward reachable sets of multi-agent systems with distributed NNC architectures. After simplifying the dynamics, we presented a method to split the overall reachability problem into multiple smaller reachability problems. We then extended this approach to account for model uncertainty. The effectiveness of this method was demonstrated on realistic examples, and it was shown to be significantly faster than the overall reachability method whilst producing the same bounds. It should also be noted that the method presented in this paper can be applied to any system that can be decomposed into the form in (\ref{eq:agentDynamics}); it is not necessarily specific to multi-agent systems.

Opportunities for future work include using the general framework presented in this paper to improve the efficiency of other reachability methods, such as LP-based methods. Also, further consideration could be given to synthesis of the NNCs, and reachability methods could be incorporated into the training process. Other types of activation functions and sets could also be considered.

%\addtolength{\textheight}{-3cm}   % This command serves to balance the column lengths
                                  % on the last page of the document manually. It shortens
                                  % the textheight of the last page by a suitable amount.
                                  % This command does not take effect until the next page
                                  % so it should come on the page before the last. Make
                                  % sure that you do not shorten the textheight too much.

% \newpage
% \begin{thebibliography}{99}
% \end{thebibliography}
\def\url#1{}
\bibliographystyle{IEEEtran}
\bibliography{mybib.bib}

\end{document}